\documentclass[english,aps,prl,twocolumn,graphics,superscriptadgraphicx,superscriptaddress]{revtex4-2}

\usepackage[utf8]{inputenc}
\usepackage{balance}

\usepackage{amsmath,mathtools}
\usepackage{amssymb}
\usepackage{graphicx}
\usepackage{verbatim}
\usepackage{braket}
\usepackage{soul}
\usepackage{bm}
\usepackage{gensymb}
\usepackage{diagbox}
\usepackage{siunitx}
\usepackage[T1]{fontenc}
\usepackage[english]{babel}
\usepackage{mathtools}
\usepackage{newtxtext}  
\usepackage{amsmath}
\usepackage{amsfonts}
\usepackage{hyperref}
\usepackage{float}

\renewcommand\vec{\boldsymbol}

\begin{document}

\title{Localized Excitons and Landau-Level Mixing in Time-Reversal Symmetric Pairs of Chern Bands}

\author{Guopeng Xu}
\affiliation{Department of Physics and Astronomy, University of Kentucky, Lexington, Kentucky 40506-0055, USA}
\author{Nemin Wei}
\affiliation{Department of Physics, Yale University, New Haven, CT 06520, USA}
\author{Inti Sodemann Villadiego}
\affiliation{Institut f\"ur Theoretische Physik, Universit\"at Leipzig, Br\"uderstra\ss e 16, 04103, Leipzig, Germany}
\author{Chunli Huang}
\affiliation{Department of Physics and Astronomy, University of Kentucky, Lexington, Kentucky 40506-0055, USA}
\date{\today} 

\begin{abstract}
We study Landau-level mixing in a time-reversal–symmetric Hamiltonian composed of two sets of Landau levels with opposite magnetic field, relevant to moiré minibands in twisted homobilayer transition-metal dichalcogenides in the adiabatic limit, where electrons in opposite valleys have flat Chern bands with opposite Chern numbers. Strong spin–orbit coupling polarizes spins in opposite directions in opposite valleys, separating Coulomb interactions into like-spin ($V^{\uparrow\uparrow}$) and opposite-spin ($V^{\uparrow\downarrow}$). Using degenerate perturbation theory, we compute Landau-level mixing corrections to $V^{\uparrow\uparrow}$ and $V^{\uparrow\downarrow}$ for different filling fractions. In the lowest Landau level, screening exhibits an even–odd effect: $V^{\uparrow\uparrow}$ is reduced more strongly than $V^{\uparrow\downarrow}$ in even-$m$ angular momentum Haldane pseudopotential and less strongly in odd-$m$ angular momentum ones. In the first Landau level, the short-range part ($m=0,1$) of $V^{\uparrow\downarrow}$ is reduced comparably to $V^{\uparrow\uparrow}$, while the strongest spin anisotropy appears in the $m=2$ pseudopotential. These novel short-range spin correlations have important implications for candidate correlated phases of fractional quantum spin Hall insulators. A distinctive feature of this time-reversal–symmetric Hamiltonian, absent in conventional quantum Hall systems, is that spin-flip excitations form localized quasiparticles. We compute their excitation spectrum and predict a non-monotonic dependence of the ordering temperature of Chern ferromagnetism in MoTe$_2$ on the Landau-level mixing parameter.
\end{abstract}


\maketitle

\textit{Introduction:} Moiré superlattices of twisted transition metal dichalcogenides (TMDs) have emerged as highly tunable platforms where many paradigmatic electronic phases can be realized—from magnets \cite{doi:10.1126/science.adg4268,Wang_2022,An_2025}, Mott insulators \cite{Tang2020,Wang2020,Shimazaki2020}, and generalized Wigner crystals \cite{Li2021,PhysRevB.103.125146,Regan2020,Xu2020,Huang2021,Jin2021,Tang2022} to superconductors \cite{Guo_2025,Xia2025,xu2025signaturesunconventionalsuperconductivitynear,https://doi.org/10.7910/dvn/y5mtri} and integer quantum anomalous Hall insulators \cite{gao2025probingquantumanomaloushall,PhysRevLett.132.146401}. 
More recently, fractional quantum anomalous Hall insulators \cite{zeng2023,Cai2023,Park2023,PhysRevX.13.031037} and fractional quantum spin Hall state \cite{Kang2024,kang2025timereversalsymmetrybreakingfractional} have also been reported in twisted homobilayer MoTe$_2$.
Extensive theoretical efforts have been devoted to understanding these fractionalized phases \cite{PhysRevResearch.3.L032070,Reddy_2023,Wang_2024,PhysRevB.109.205121,PhysRevResearch.7.023083,PhysRevResearch.5.L032022,PhysRevLett.131.136502}, often relying on large-scale numerical simulations where Coulomb or other density–density interactions are projected into the relevant Chern bands. In realistic moiré TMDs, however, the Coulomb interaction typically exceeds the moir\'e minibandwidth, so projection alone is insufficient and tends to yield spin-isotropic interactions that favor simple Chern ferromagnet at odd filling fractions. Such Chern ferromagnets would produce a quantum anomalous Hall effect, which is not the state observed at moiré hole filling $\nu=3$ in $2.1^\circ$ twisted MoTe$_2$ \cite{Kang2024,kang2025timereversalsymmetrybreakingfractional}.
Ref.~\cite{PhysRevResearch.7.023083} noted that virtual mixing with higher moiré minibands can generate spin-anisotropic interactions, and modeled these effects by treating spin anisotropy as equivalent to layer anisotropy.
However, a systematic account of band-mixing processes has been lacking. In this Letter, we close this gap by deriving effective interaction parameters for a minimal model of twisted TMDs and providing experimentally testable predictions that follow directly from them.

\textit{Model:} 
The $K$-valley twisted homobilayer TMDs can host narrow Chern bands over a range of small twist angles \cite{PhysRevLett.122.086402,devakul2021magic,zhang2024polarization}. Within the adiabatic model for these bilayer TMDs \cite{PhysRevLett.132.096602,PhysRevB.110.035130}, electrons in $\pm K$ valleys experience opposite emergent magnetic fields generated by the real-space layer-pseudospin skyrmion texture \cite{PhysRevLett.122.086402}. The magnitude $B_{\text{sky}}$ of the average magnetic field equals one flux quanta per moir\'e unit cell in individual valley, yielding magnetic quasi-Bloch bands that correspond to the Chern bands. This observation motivates a minimal description of the time-reversal pairs of narrow Chern bands in terms of Landau levels in uniform magnetic fields of opposite signs. The same approach was taken previously in Ref.~\cite{zou2025valley, PhysRevLett.124.166601}.
Owing to strong spin–orbit coupling, spin and valley degrees of freedom are locked. Within the adiabatic approximation, the minimal description of TMDs can be captured by the Hamiltonian$H = T + V$ :
\begin{align}
   T=\sum_{nm }\left( n c^\dagger_{nm \uparrow }c_{nm \uparrow}  +n c^\dagger_{\bar{n}\bar{m} \downarrow  }c_{\bar{n} \bar{m}\downarrow} \right) \label{eq:T}
\end{align}
\begin{align}
    V= \frac{\kappa}{2}\sum_{12'34'}\sum_{\sigma\sigma'}
    \bra{1\sigma,2'\sigma'}V_c\ket{3\sigma,4'\sigma'}
    c_{1\sigma}^{\dagger} c_{2'\sigma'}^{\dagger} c_{4'\sigma'} c_{3\sigma}
    \label{eq:Coulomb_V_second}
\end{align}
 We measure all energies in units of the cyclotron energy $\hbar\omega_{c}$ where $\omega_{c}=eB_{\mathrm{sky}}/(m^{\ast}c)$ and use composite indices $1 \equiv (n_1 m_1)$ to label the Landau-level (LL) and guiding-center quantum numbers. Wavefunctions with opposite spin-projection have opposite chirality, $\phi_{nm}(r) = \langle r | c_{nm\uparrow}^\dagger | 0 \rangle$, $\phi_{\bar{n}\bar{m}}(r) = \langle r | c_{\bar{n}\bar{m}\downarrow}^\dagger | 0 \rangle$, with $\phi^*_{\bar{n}\bar{m}}(r) = \phi_{nm}(r)$ by time-reversal symmetry. We used overbar denotes the state of opposite chirality.
 In $V$, $V_c=1/|r_1-r_2|$ and for like-spin interaction ($\sigma'=\sigma$), the composite indices carry the same chirality ($2'=2$ and $4'=4$), while for opposite spins interaction ($\sigma'=-\sigma$), they switch chirality ($2'=\bar 2$ and $4'=\bar 4$). They satisfy the following relations
\begin{equation}
V^{\sigma \bar\sigma}_{1\bar2,3\bar4} = V^{\sigma \bar\sigma}_{14,32}\;,\; V^{\sigma \bar\sigma}_{14,32}=V^{\sigma \sigma}_{14,32},
\end{equation}
where the first relation follows from taking the time-reversal conjugate of the wavefunction and holds for any $V$. The second relation reflects the spin-independence of the bare Coulomb potential but breaks down once Landau-level mixing is included.
One advantage of this Landau-level description is that the relative importance of interaction and band dispersion is controlled by a \emph{single} dimensionless coupling constant \begin{equation}
    \kappa = \frac{e^{2}}{\epsilon \ell_{B} \hbar\omega_{c}} 
\end{equation}
For twisted MoTe$_2$ encapsulated by hBN substrate, this is approximately $\kappa\sim 6.5$ at twisted angle $\theta\approx2.1^\circ$ \cite{PhysRevLett.132.096602}. Similar models have been studied in previous works~\cite{kwan2021exciton,kwan2022excitonic,kwan2025textured,sodemann2024halperin}, but without accounting for LL mixing.

\textit{Landau-level projection:} 
A standard route to capture strong correlations in flat bands is to project the full Hamiltonian $H$ onto a chosen flatband, where the kinetic energy is quenched. 
The $n$-th LL projected Hamiltonian is
\begin{equation}
P_n VP_n=\frac{1}{2A}\sum_{\mathbf q\neq0} \kappa V_{n}(q)\,\bar \rho^\dagger(\mathbf q)\bar \rho(\mathbf q),
\end{equation}
with $P_n$ the projector onto the $n$-th LL, i.e. the many-body degenerate Hilbert space defined by the eigenvalues of $T$ in Eq.~\ref{eq:T}. Here $\bar \rho(\mathbf q)=\bar \rho_\uparrow(\mathbf q)+\bar \rho_\downarrow(\mathbf q)$ is the LL-projected density operator and 
$V_{n}(q)=V(q)|F_{nn}(q)|^2$ the LL-projected Coulomb potential, with $V(q)=2\pi /q$. The form factor $F_{nn}(q)$ reflects the smearing of the electron coordinate over the magnetic length but is independent of electron chirality. Since $V_{n}(q)\geq 0$, the energy expectation is also non-negative,
\begin{equation} \label{eq:E>0}
\langle \Psi|P_nVP_n|\Psi\rangle\geq 0.
\end{equation}
At half filling of the two-fold spin-degenerate LL, the exact ground state is the quantum Hall ferromagnet (QHF),
$|\text{QHF}\rangle=\prod_k c^\dagger_{k\uparrow}|0\rangle$, 
in which the majority spin LL is fully occupied and the minority spin LL is empty. 
Since no density-fluctuations can be generated in this state, $\bar \rho(\mathbf q)|\text{QHF}\rangle=0$,  the QHF is the exact ground state of the LL-projected Hamiltonian at half filling, and any instability must arise from LL mixing.

\begin{figure}[t!]
    \centering
    \includegraphics[width=1.0\linewidth]{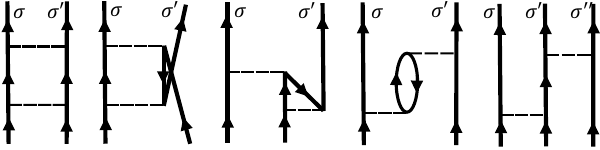}
    \caption{Landau-level–mixing corrections to the Coulomb potential. From left to right: two-body terms (BCS, double-exchange, vertex, bubble) and the  three-body term.
    } 
    \label{fig:one_loop_diagram}
\end{figure}


\textit{Landau-level mixing and Haldane Pseudopotentials:} The LL-projected Hamiltonian to second order in $\kappa$ is \cite{PhysRevB.87.245425,PhysRevLett.134.046501}:
\begin{align}
\mathcal{H}^{\text{eff}}_n &=P_n  V P_n - P_n V P_\perp \frac{1}{ T -E_0} P_\perp V P_n \notag \\
&\equiv \kappa V_n + \kappa^2 \delta V_n + \kappa^2 W_n\end{align}
where $P_\perp = 1 - P_n$ and $E_0$ is the eigenvalue of $T$ defining the n-th LL manifold. The effective Hamiltonian $\mathcal{H}^{\text{eff}}_n$ generate interaction up to three-body and is independent of the filling fraction of the $n$-th LL. The three-body contribution, $W_n$, will be discussed in the next section; here we focus on the two-body sector $V_n$ and $\delta V_n$.  The two-body correction, $\delta V_n$, can be decomposed into four distinct contributions, corresponding to the Feynman diagrams in Fig.~\ref{fig:one_loop_diagram},  
\begin{align}
\delta V_n = \sum_{\alpha=\text{BCS, EX, BB, VC}} \delta V^{(\alpha)}_n.
\end{align}

A convenient basis for comparing the strength of the $n$-th LL projected two-body interaction is the center-of-mass and relative guiding-center states $\ket{m,M}$:  
\begin{align}
    &\langle 1\uparrow,2\uparrow|V^{\uparrow\uparrow}|3\uparrow,4\uparrow\rangle \notag \\
    &= \sum_{m,M}V_{n,m}^{\uparrow\uparrow} \langle m_1,m_2|M,m\rangle\langle M,m|m_3,m_4\rangle \\
    &\langle 1\uparrow,\bar2\downarrow|V^{\uparrow\downarrow}|3\uparrow,\bar4\downarrow\rangle =\langle1\uparrow,4\downarrow|V^{\uparrow\downarrow}|3\uparrow,2\downarrow\rangle \notag\\
    &= \sum_{m,M}V_{n,m}^{\uparrow\downarrow}\langle m_1,m_4|M,m\rangle\langle M,m|m_3,m_2\rangle
\end{align}
where $\langle m_1,m_2|M,m\rangle = R^{M+m}_{m,m_1}\,\delta_{m_1+m_2,M+m}$ is the standard transformation between the single-particle and COM–relative bases \cite{supmat}, and $V_{n,m}^{\sigma\sigma'}$ are the Haldane pseudopotentials in the $n$-th LL. For the opposite-spin interaction, we swap $2$ and $4$ (removing the bar) to implement a particle–hole transformation on one of the spin species, so that electron and hole expereince the same effective magnetic. The same COM–relative basis applies to the second-order corrections $\delta V_n$ as well.

\begin{table*}[t!]
    \centering
    \renewcommand{\arraystretch}{1.2}
    \begin{tabular}{|c|cc|cc|cc|cc|cc|c|c|}
        \hline
        Haldane Pseudopotentials & \multicolumn{2}{c|}{$m=0$} & \multicolumn{2}{c|}{$m=1$} & \multicolumn{2}{c|}{$m=2$} & \multicolumn{2}{c|}{$m=3$} & \multicolumn{2}{c|}{$m=4$} & \multicolumn{2}{c|}{Fock Self Energy $\Sigma_F^{(\alpha)}$} \\
        \cline{2-13}
        & $\uparrow\uparrow$ & $\uparrow\downarrow$ & $\uparrow\uparrow$ & $\uparrow\downarrow$
        & $\uparrow\uparrow$ & $\uparrow\downarrow$ & $\uparrow\uparrow$ & $\uparrow\downarrow$
        & $\uparrow\uparrow$ & $\uparrow\downarrow$ & Bare Value &  Correction     \\
        \hline\hline
        \multicolumn{13}{|c|}{\textbf{Lowest Landau level (\(n=0\))}} \\
        \hline
        $V_{0,m}$                & \multicolumn{2}{c|}{0.8862} & \multicolumn{2}{c|}{0.4431}
                                 & \multicolumn{2}{c|}{0.3323} & \multicolumn{2}{c|}{0.2769}
                                 & \multicolumn{2}{c|}{0.2423} & $\sqrt{\frac{\pi}{2}}$ & -0.6410 \\
        $\delta V^{\mathrm{BCS}}_{0,m}$
                                 & -0.3457 & -0.2837 & -0.0328 & -0.0997
                                 & -0.0112 &  0.0036 & -0.0055 & -0.0118
                                 & -0.0033 & -0.0023 &   &  \\
        \hline\hline 
        \multicolumn{13}{|c|}{\textbf{First Landau level (\(n=1\))}} \\
        \hline
        $V_{n=1,m}$              & \multicolumn{2}{c|}{0.6093} & \multicolumn{2}{c|}{0.4154}
                                 & \multicolumn{2}{c|}{0.4500} & \multicolumn{2}{c|}{0.3150}
                                 & \multicolumn{2}{c|}{0.2635} &  &  \\
        $\delta V^{\mathrm{BB}}_{n=1,m}$
                                 & -0.3930 & -0.3930 & -0.2038 & -0.2038
                                 & -0.1981 & -0.1981 & -0.1119 & -0.1119
                                 & -0.0535 & -0.0535 &  & -0.6244 \\
        $\delta V^{\mathrm{VC}}_{n=1,m}$
                                 &  0.0750 &  0.0750 & -0.0475 & -0.0475
                                 &  0.0870 &  0.0870 &  0.0141 &  0.0141
                                 & -0.0128 & -0.0128 &  & 0.3878 \\
        $\delta V^{\mathrm{EX}}_{n=1,m}$
                                 &  0.0247 &  0.0750 &  0.0706 &  0.0135
                                 &  0.0803 &  0.0841 &  0.0186 &  0.0152
                                 & -0.0031 &  0.0066 & $\frac{3}{4}\sqrt{\frac{\pi}{2}}$ & -0.0002 \\
        $\delta V^{\mathrm{BCS}}_{n=1,m}$
                                 & -0.0890 & -0.1550 & -0.0347 &  0.0015
                                 & -0.1498 & -0.1091 & -0.0264 & -0.0329
                                 & -0.0116 & -0.0151 &  & -0.3688 \\
        $\sum_{\alpha}\delta V^{\alpha}_{n=1,m}$
                                 & -0.3823 & -0.3980 & -0.2154 & -0.2363
                                 & -0.1806 & -0.1361 & -0.1056 & -0.1155
                                 & -0.0810 & -0.0748 &  & -0.6056 \\
        \hline
    \end{tabular}
    \caption{Spin-resolved Haldane pseudopotentials in the lowest (\(n=0\)) and first (\(n=1\)) Landau levels. Each \(m\)-sector has two columns corresponding to like-spin (\(\uparrow\uparrow\)) and opposite-spin (\(\uparrow\downarrow\)) Coulomb interactions. For each Landau level the first row gives the bare projected interaction, followed by second-order corrections from the indicated diagrammatic processes. In the last column, we show the second order correction to the self energy arising from those diagrams.}
    \label{tab:haldane_pseudopotentials_combined}
\end{table*}

Table~\ref{tab:haldane_pseudopotentials_combined} lists the matrix elements of $\mathcal{H}^{\text{eff}}_n$ for $n=0,1$ and relative angular momentum $m=1,\dots,4$. The $n=0$ case is simpler, as it reduces to a two-electron problem. At first order in $\kappa$, the 0LL-projected interaction is spin independent, $V^{\sigma\sigma}_{n=0,m}=V^{\sigma\bar\sigma}_{n=0,m}$. For the correction $\delta V_{n=0,m}$, only the BCS diagram contributes, since no lower LLs are occupied.  BCS diagram  leads to an even–odd effect: equal-spin interaction is reduced more strongly than opposite-spin interaction in even-$m$ channels, and less strongly in odd-$m$ channels. This behavior originates from matrix-element effects \cite{supmat}, as the phase space for particle–particle excitations is identical for both like-spin and opposite-spin. While the BCS correction is always negative in the like-spin channel, it is not guaranteed to be negative in the opposite-spin channel and can even become positive, e.g., $\delta V^{\text{BCS},\uparrow\downarrow}_{n=0,m=2}$.

The situation in the $n=1$ LL is more complicated, since the fully occupied $n=0$ LL with both spin projections provides phase space for particle–hole excitations. An interesting feature is the violation of the Galitskii rule \cite{Galitskii1958}. In ordinary dilute Fermi gases or in the Hubbard model \cite{kanamori1963electron}, Galitskii’s rule states that the particle–particle (BCS) diagram dominates due to its larger phase space. However, for electrons in a magnetic field, this does not apply: although the BCS diagram still has the largest phase space, most virtual particle–particle processes are strongly suppressed by matrix-element constraints \cite{supmat}. Instead, the bubble (BB) particle–hole diagram, which screens the total charge density, provides the dominant reduction of the pseudopotentials. This property already exists in earlier Landau-level mixing data for electron-gas in magnetic field\cite{PhysRevB.87.245129,PhysRevB.87.155426,PhysRevB.87.245425} but was not emphasized explicitly.
As in the $n=0$ case, the $n=1$ projected Coulomb potential is spin independent, with the pseudopotential exhibiting the well-known local minimum at $m=1$. Spin anisotropy arises only from the EX and BCS diagrams. A detailed analysis of the contributions from different virtual excitations is given in the appendix \cite{supmat}; here we summarize the main results. Opposite-spin interactions are screened slightly more than like-spin interactions in both the $m=0$ and $m=1$ channels, a counterintuitive outcome. The largest spin anisotropy appears at $m=2$, where like-spin interactions are screened more strongly than opposite-spin interactions. Notably, for moderate coupling constant $\kappa>0.62$, \cite{supmat} the opposite-spin pseudopotential is largest at $m=2$, and this has non-trivial consequence for the excited state spectrum as we now show.

\begin{table}[b]
\centering
\caption{Three-body interaction contribution to the exciton energy}
\begin{tabular}{c c c c c c}
\hline
$\kappa^2$ & $m=0$ & $m=1$ & $m=2$ & $m=3$ & $m=4$ \\
\hline
$W_{n=0,m}$   & 0.0358 & 0.0972   & 0.0937   & 0.0587   & 0.0587 \\
$W_{n=1,m}$   & -0.0011   & 0.022   & 0.0048  & -0.0384   & -0.0333 \\
\hline
\end{tabular}\label{tab:three_b_contribution}
\end{table}

\textit{Localized excitons:} Including LL mixing introduces spin anisotropy ($V^{\uparrow\uparrow}\neq V^{\uparrow\downarrow}$), and $V^{\sigma\sigma'}_n(q)$ need not remain positive definite. Consequently, the expectation value of the effective Hamiltonian $\mathcal{H}^{\text{eff}}_n$ to second order in $\kappa$ is no longer guaranteed to be non-negative. Thus, the $\ket{\text{QHF}}$ with maximum spin-polarization $S_z=N\hbar/2$ is no longer guaranteed to be the ground state at half filling of a two-fold spin-degenerate LL. A natural approach to investigate this instability is to study the $S_z=(N-2)\hbar/2$ sector, corresponding to a single spin exciton. A distinctive feature of our time-reversal–symmetric flatband Hamiltonian is that the spin-flip particle–hole excitations form localized quasiparticles. As discussed in previous works \cite{kwan2021exciton,stefanidis2020excitonic}, this localization arises because an electron of one spin and a hole of the opposite spin experience the same effective magnetic field, binding into a neutral exciton. This sharply contrasts with conventional QHF, where spin-flip excitations are running waves \cite{kallin1984excitations}. Thus, while the ground states of the two systems may look similar, their finite-temperature behavior is qualitatively different.  

The many-body spectrum in the $S_z=(N/2-1)\hbar$ sector can be obtained exactly using the equation of motion method, due to the restricted phase space for density fluctuations in a projected LL. As shown in Ref.~\cite{supmat}, the exciton energy in the $n$-th LL is
\begin{align}\label{eq:g RPA}
    A_{n,m}=\Sigma^F_n - \mathcal{V}^{\uparrow\downarrow}_{n,m} + W_{n,m},
\end{align}
where the energy depends only on the relative guiding center $m$. For each $m$, the states are macroscopically degenerate in the COM guiding center, forming a Landau-level of excitons \cite{kwan2021exciton}. Here $\Sigma^F_n=\sum_{m} 2(-1)^m\,\mathcal{V}_{nm}^{\uparrow\uparrow}$
is the Fock self-energy, where $\mathcal{V}^{\uparrow\uparrow}_{n,m}=\kappa V_{nm} + \kappa^2 \delta V^{\uparrow\uparrow}_{nm}$ includes leading LL-mixing corrections (values given in the last two columns of Table~\ref{tab:haldane_pseudopotentials_combined}). $\Sigma^F$ is the energy cost to remove a majority spin from a uniform QHF and is thus
independent of $m$. The second term, $\mathcal{V}^{\uparrow\downarrow}_{n,m}=\kappa V^{\uparrow\downarrow}_{n,m}+\kappa^2 \delta V^{\uparrow\downarrow}_{n,m}$, describes the electron–hole attraction and competes with the Fock self-energy. Finally, $W_{n,m}$ is the three-body contribution arising from normal ordering with respect to the half-filled LL (Table~\ref{tab:three_b_contribution}), which is small compared to the other two terms. Together, these contributions determine the exciton energy exactly to order $\kappa^2$.  

\begin{figure*}[t!]
     \centering
    \includegraphics[width=1\linewidth]{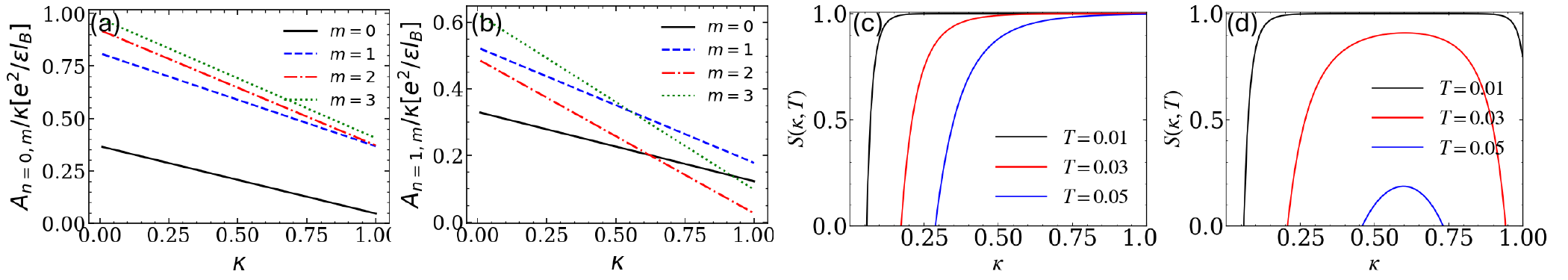}
    \caption{(a-b)Spin–exciton energy(divided by $\kappa$) vs $\kappa$ in the $n=0$ and $n=1$ Landau level. An exciton energy level crossing occurs between the $m=0$ and $m=2$ mode in $n=1$. (c–d) interaction strength dependence of the spin polarization (order parameter) for the $\nu=3$ quantum Hall ferromagnet (half-filled $n=1$ LL) at different temperature, computed using the bare Coulomb interaction (c) and the renormalized Coulomb interaction (d). The temperautre is in unit of $\hbar\omega_c/k_B$.}
    \label{fig:exciton_energys}
\end{figure*}

Fig.~\ref{fig:exciton_energys}(a,b) show $A_{n,m}/\kappa$ vs $\kappa$. At small $\kappa$, QHF is the exact ground state [Eq.~\eqref{eq:E>0}], implying that the exchange cost exceeds the exciton attraction.
With increasing $\kappa$, LL mixing becomes more important and the exciton energy softens due to stronger reduction of the exchange energy relative to the opposite-spin Haldane pseudopotential. The nature of this softening differs qualitatively between $n=0$ and $n=1$ LLs. In the $n=0$ LL [Fig.~\ref{fig:exciton_energys}(a)], it occurs for excitons with angular momentum $m=0$, where the exciton attraction is always the strongest. In the $n=1$ case [Fig.~\ref{fig:exciton_energys}(b)], there is an exciton energy level crossing between the $m=2$ and $m=0$ angular momentum states near $\kappa \approx 0.6$. This arises from the combined effect of stronger screening in the $m=0$ channel and the peculiar structure of the $n=1$ Haldane pseudopotential, which exhibits a local minimum at $m=1$. 
 After the level crossing, the $m=2$ exciton continues to soften,
eventually crossing zero, indicating the appearance of a non-trivial state. 

One important consequence of LL mixing is that the spin gap, i.e.~the minimum of single spin-reversal excitation, $\min[A_{n,m}]$, becomes a non-monotonic function of $\kappa$. This non-monotonicity may imprint itself as a non-monotonic dependence of magnetization vs $\kappa$. Within spin-wave theory, the spin polarization (order parameter) of the QHF in the $n$-th LL can be estimated as  
\begin{align}
    S(\kappa,T) = 1 - 2\sum_m \frac{1}{e^{A_{n,m}/k_BT}-1}.
\end{align}
The above expression is in unit of $\hbar/2$. Figures~\ref{fig:exciton_energys}(c,d) show $S(\kappa,T)$ for the $n=1$ QHF with and without LL mixing. In the absence of LL mixing [Fig.~\ref{fig:exciton_energys}(c)], $S(\kappa,T)$ increases monotonically with $\kappa$ as the exciton energy hardens. By contrast, when LL mixing is included [Fig.~\ref{fig:exciton_energys}(d)], $S(\kappa,T)$ becomes a non-monotonic function of $\kappa$, with mode softening shifting the magnetization curves to lower temperatures. The $n=0$ QHF exhibits the same qualitative behavior and is therefore not shown.  Since magnetization is a routinely measurable quantity in moir\'e semiconductors  with magnetic circular dichroism \cite{Li_2021,Xu_2022,Wang_2022,Huang_2017,Anderson_2023} and nano-SQUID \cite{redekop2024direct,li2025universal}, and spin  polarization is part of the total magnetization, these predictions provide an experimental test of LL-mixing effects and their role in destabilizing the QHF.

\textit{Summary and Discussions:} 
We have analyzed Landau-level mixing in a time-reversal–symmetric flatband Hamiltonian relevant to twisted homobilayer TMDs. We derived effective interactions up to second order in $\kappa$ and demonstrated that LL mixing generates spin-anisotropic corrections to the Haldane pseudopotentials. In the lowest LL, screening exhibits an even–odd effect, while in the first LL the strongest anisotropy occurs in the $m=2$ channel. A distinctive property of this Hamiltonian is that spin-flip excitations form \emph{localized excitons}, in sharp contrast to the delocalized spin waves of conventional QHF. We computed their excitation spectrum exactly to order $\kappa^2$, showing that LL mixing induces mode softening: in $n=0$ the lowest mode lies at $m=0$, while in $n=1$ a level crossing shifts the softest mode to $m=2$. This softening directly controls the QHF magnetization and produces a non-monotonic dependence of the ordering temperature on $\kappa$. Our results demonstrate a concrete mechanism for how Chern ferromagnetism in moiré TMDs can become unstable to exciton proliferation, with the $n=1$ state likely favoring excitons carrying finite angular momentum $m=2$.  

Determining the states that are stabilized at finite exciton density is non-trivial. One natural possibility is that excitons Bose condense, leading to valley-symmetry–broken states. 
However, since spin-flip excitons behave as bosons in a magnetic field, another interesting possibility is the formation of correlated phases analogous to Bosonic Laughlin states \cite{kwan2022excitonic,stefanidis2020excitonic}.
These studies have not considered, however, the phases that would be realized if the proliferating excitons carry finite angular momentum, which appears to be a natural possibility in the $n=1$ LL as we have shown. This question is especially pressing because this LL has been argued to play a key role in understanding the physics at filling $\nu=3$, where experiments have reported signatures of a fractional quantum spin Hall state. While several theoretical proposals have been made for this state\cite{PhysRevResearch.7.023083,PhysRevX.15.021063,PhysRevLett.133.146503,PhysRevB.110.045114,PhysRevB.110.155102,zhang2024vortexspinliquidfractional}, detailed  microscopic pictures for why they would be energetically favorable are still lacking. Our study thus provides a concrete model for modified interactions by Landau level mixing that can serve as the starting point of numerical studies to elucidate the nature of the phases that compete with the Chern magnet at this filling.

\textit{Acknowledgments}: We acknowledge insightful discussion with Ahmed Abouelkomsan and 
Ajit Balram. 
We acknowledge support from the Deutsche Forschungs-gemeinschaft (DFG) through research grants Project No.542614019, No.518372354, No.555335098. 
\bibliographystyle{ieeetr}

\bibliography{references}
\newpage

\clearpage
\widetext

\begin{center}
\textbf{\large Supplementary Material: Localized Excitons and Landau-Level Mixing in Time-Reversal Symmetric Pairs of Chern Bands}
\end{center}

\setcounter{equation}{0}
\setcounter{figure}{0}
\setcounter{table}{0}
\setcounter{page}{1}
\makeatletter
\renewcommand{\theequation}{S\arabic{equation}}
\renewcommand{\thefigure}{S\arabic{figure}}

\newcounter{proplabel}



\tableofcontents

\section{Adiabatic Model of twisted TMDs}

In this section, we briefly talk about the adiabatic model of twisted TMDs,  the details can be found in Ref.~\cite{PhysRevLett.132.096602}. Its explicit form is given by the following
\begin{align}
    h_\sigma = \frac{1}{2m}(p+ \text{sgn}(\sigma) \frac{eA}{c})^2,
\end{align}
where the emergent gauge field $A$ share the same periodicity as the moir\'e potential, leading to emergent effective magnetic field $B=\nabla \times A = \frac{\hbar }{2e} \vec n\cdot (\partial_x \vec n \times \partial_y \vec n)$ where $\vec n$ is the spin texture of the Skyrmion lattices\cite{PhysRevLett.122.086402}.The averaged effective magnetic field, denoted $B_{\text{sky}}$, contains one 
flux quantum per moir\'e unit cell. For simplicity, we retain only the 
homogeneous component and choose the symmetric gauge, 
$\mathbf{A}=\tfrac{B_{\text{sky}}}{2}(-y,x,0)$. The index $\sigma$ labels 
spin (or valley, owing to strong spin--valley locking in twisted TMDs,
we will refer to it as spin).
 The  single particle Hamiltonian for spin-up particles is related to the spin-down by time reversal symmetry:

\begin{align}
    h_{\downarrow} = h^*_{\uparrow},
\end{align}

so the corresponding orbital Landau-level wavefunction also satisfies:

\begin{align}\label{eq:wavefunction property}
\phi_{nm\downarrow} = \phi_{nm\uparrow}^*.
\end{align}

To avoid confusion, we will denote the spin-up LL wavefunction as $\phi_{nm\uparrow} \equiv \phi_{nm}$ and spin-down LL wavefunction as $\phi_{nm\downarrow} = \phi_{\bar n \bar m}$, the bar over head indicate it experience opposite direction of magnetic field as the spin-up states. The LL wavefunction in symmetric gauge is given by~\cite{macdonald1994introductionphysicsquantumhall}:

\begin{align}
    \phi_{nm} &= \frac{1}{\sqrt{2\pi}}F_{mn}(i\bar z) \label{eq:wavefunction_symmetric_gauge}\\
    F_{n'n}(z)& = \sqrt{\frac{\min(n',n)!}{\max(n',n)!}}\left(-i\frac{|z|}{\sqrt{2}}\right)^{|n'-n|}e^{i\theta(n'-n)}L^{|n'-n|}_{\min(n',n)} \left(\frac{|z|^2}{2}\right) e^{-\frac{|z|^2}{4}} \label{eq:form_factor}
\end{align}
where $z=x+iy=|z|e^{i\theta}$ and $F_{n'n}$ is known as form factor. The position is in unit of magnetic length of the emergent magnetic field and we set it to be $1$. The many-body Hamiltonian is given by:

\begin{align}
    H = \sum_{\sigma}\int d^2r \psi^\dagger_{\sigma}(r)h_\sigma(r) \psi_{\sigma}(r)+\sum_{\sigma,\sigma'}\frac{1}{2}\int\int d^2rd^2r':\psi_{\sigma}^\dagger(r)\psi_{\sigma}(r)\frac{e^2}{\epsilon|r-r'|}\psi^\dagger_{\sigma'}(r')\psi_{\sigma'}(r'):
\end{align}
The field operators can be expanded in the Landau-level basis as
\[
\psi_{\uparrow} = \sum_{nm}\phi_{nm}c_{nm\uparrow}, 
\quad 
\psi_{\downarrow} = \sum_{nm}\phi_{\bar{n}\bar{m}}c_{nm\downarrow}.
\]
The corresponding many-body Hamiltonian, expressed in units of the cyclotron 
energy, is
\begin{align}
H = \sum_{nm} n \left(c^\dagger_{nm\uparrow}c_{nm\uparrow} 
+ c^\dagger_{\bar n \bar m \downarrow}c_{nm\downarrow}\right)+ \frac{\kappa}{2} \sum_{1234,\sigma} 
\Big( V_{12,34}^{\uparrow\uparrow} \, c^{\dagger}_{1\sigma} c^\dagger_{2\sigma} 
c_{4\sigma} c_{3\sigma} 
+ V_{1\bar{2},3\bar{4}}^{\uparrow\downarrow} \, c^{\dagger}_{1\sigma} 
c^\dagger_{2\bar\sigma} c_{4\bar\sigma} c_{3\sigma} \Big).
\label{eq:Hamiltonian}
\end{align}
where the dimensioinless parameter $\kappa$ is defined as the ratio of the Coulomb interaction energy to the Landau-level kinetic energy:
\begin{align}
    \kappa = \frac{e^2}{\epsilon \hbar \omega_c l_B}
\end{align}
and the Coulomb interaction matrix elements are defined as:
\begin{align}
    V^{\uparrow\uparrow}_{12,34}
    & = \int \int d^2r d^2r' \phi_{n_1m_1}^*(r)\phi_{n_2m_2}^*(r')\frac{1}{|r-r'|}\phi_{n_3m_3}(r)\phi_{n_4m_4}(r') \label{eq:like-spin matrix elements} \\
        V^{\uparrow\downarrow}_{1\bar{2
    },3\bar{4}}
    & = \int \int d^2r d^2r' \phi_{n_1m_1}^*(r)\phi_{\bar{n}_2\bar{m}_2}^*(r')\frac{1}{|r-r'|}\phi_{n_3m_3}(r)\phi_{\bar{n}_4\bar{m}_4}(r')\label{eq:opposite-spin matrix elements}
\end{align}

The index ``1'' is a shorthand for $n_1m_1$. Same-spin interactions are 
denoted by the superscript $\uparrow\uparrow$, and opposite-spin interactions 
by $\uparrow\downarrow$. In the symmetric gauge, the Coulomb matrxi elements are real, namely,
$V_{12,34}=V_{34,12}$, this equation hold no matter its same-spin or opposite-spin.

Using Eq.~\eqref{eq:wavefunction property}, bare Coulomb matrix elements satisfies 
\begin{align}
V^{\uparrow\downarrow}_{1\bar{2
    },3\bar{4}}=V^{\uparrow\downarrow}_{14,32}= V^{\uparrow\uparrow}_{14,32} \label{eq:Coulomb_symmetric}
\end{align}
The first relation follows from the chirality of the wavefunction and remains 
valid for effective Coulomb interactions, whereas the second holds only for 
the spin-independent bare Coulomb interaction.

Since the Coulomb interaction conserves angular momentum, two particles with 
the same spin (and hence the same chirality) yield nonzero matrix elements 
only if  
\begin{align}\label{eq:like-spin angular momentum}
    m_1-n_1+m_2-n_2 = m_3-n_3+m_4-n_4, 
\quad \text{(Angular momentum Conservation---like-spin particles).}
\end{align}
For two particles with opposite spins, their chiralities differ, and angular 
momentum conservation instead requires  
\begin{align}\label{eq:opposite-spin angular momentum}
    m_1-n_1-(m_2-n_2) = m_3-n_3-(m_4-n_4), 
\quad \text{(Angular momentum Conservation---opposite-spin particles).}
\end{align}
This chirality-induced modification of the angular momentum conservation law 
is the key mechanism underlying the spin anisotropy generated by Landau-level 
mixing.

Finally, we present the closed form for bare Coulomb matrix elements defined in Eq.~\eqref{eq:like-spin matrix elements} and ~\eqref{eq:opposite-spin matrix elements}. In view of the Eq.~\eqref{eq:Coulomb_symmetric}, it suffices to compute the like-spin bare Coulomb matrix elements.
The like-spin bare Coulomb matrix 
elements are most conveniently evaluated in the center-of-mass (COM) and 
relative basis, since the interaction depends only on the relative coordinate. 
Because the two particles experience the same magnetic field, 
the transformation between the single-particle basis and the COM--relative 
basis is standard and applies both in Landau-level space and in guiding-center 
space:

\begin{align}
    |m_1,m_2\rangle &= \sum_{M,m}{}_r\langle M,m|m_1,m_2\rangle|M\rangle_c|m\rangle_r \label{eq:COM basis transformation}\\
    {}_c\langle M|{}_r\langle m|m_1,m_2\rangle& = R^{m_1+m_2}_{m_2,m}\delta_{m_1+m_2,M+m}
\end{align}
the subscript $c$ and $r$ indicate COM and relative state .
The unitrary transformation matrix $R^{L}_{m,n}$ is given by 
\begin{align}\label{eq:R_function}
  R^{L}_{m,n}=  \sqrt{\frac{\binom{L}{m}}{\binom{L}{n}2^L}}\sum_{u=\max(0,m+n-L)}^{\min(m,n)} \binom{L-m}{n-u}\binom{m}{u}(-1)^u
\end{align} 
Since it is real, the inverse transformation is given by the same expression. 
The like-spin Coulomb interaction matrix elements in the COM and relative basis is given by\cite{PhysRevB.87.245425}:

\begin{align}
    {}_c\langle NM|{}_r\langle nm|V^{\uparrow\uparrow} |N'M'\rangle_c|n'm'\rangle_r &= {}_r\langle nm|V^{\uparrow\uparrow} |n'm'\rangle_r \delta_{NN'}\delta_{MM'}\delta_{n-m,n'-m'}\\
    {}_r\langle nm|V^{\uparrow\uparrow} |n'm'\rangle_r &= \frac{\Gamma(|j|+\frac{1}{2})\Gamma(l'+\frac{1}{2})}{2|j|!} \sqrt{\frac{(l+|j|)!}{\pi (l'+|j|)!l'!l!}} {}_3F_2\left(
\begin{array}{c}
 -l,\ |j|+\tfrac{1}{2},\ \tfrac{1}{2} \\
 |j|+1,\ \tfrac{1}{2}-l'
\end{array}
; 1 \right)
\end{align}
where $m,n,m',n' \geq0$  and $j=m-n$, $l=n+(j-|j|)/2$, $l'=n'+(j-|j|)/2$. For a Landau-level--projected system, the standard way to represent the Coulomb matrix elements is through the Haldane pseudopotential decomposition:
\begin{align}
    \langle 1\uparrow,2\uparrow|V^{\uparrow\uparrow}|3\uparrow,4\uparrow\rangle 
    &= \sum_{m,M} V^{\uparrow\uparrow}_{n,m}\,\langle m_1,m_2|M\rangle_c|m\rangle_r {}_c\langle M|{}_r\langle m|m_3,m_4\rangle \\
    &= \sum_{m,M} V^{\uparrow\uparrow}_{n,m}\,
    R^{M+m}_{m,m_4} R^{M+m}_{m,m_2}\,
    \delta_{m_1+m_2,M+m}\,\delta_{m_3+m_4,M+m}, \\[6pt]
    \langle 1\uparrow,\bar{2}\downarrow|V^{\uparrow\downarrow}|3\uparrow,\bar{4}\downarrow\rangle 
    &= \langle 1\uparrow,4\downarrow|V^{\uparrow\downarrow}|3\uparrow,2\downarrow\rangle \\
    &= \sum_{m,M} V^{\uparrow\downarrow}_{n,m}\,\langle m_1,m_4|M\rangle_c|m\rangle_r {}_c\langle M|{}_r\langle m|m_3,m_2\rangle \\
    &= \sum_{m,M} V^{\uparrow\downarrow}_{n,m}\,
    R^{M+m}_{m,m_4} R^{M+m}_{m,m_2}\,
    \delta_{m_1+m_4,M+m}\,\delta_{m_3+m_2,M+m}. \label{eq:opposite_spin_haldane}
\end{align}

Here, $V^{\uparrow\uparrow}_{n,m}$ denotes the like-spin (particle--particle) Haldane pseudopotential, while $V^{\uparrow\downarrow}_{n,m}$ represents the opposite-spin (particle--hole) pseudopotential, the reason we also call it particle--hole pseudopotential is because we swap between a ket state and bra state.
The pseudopotentials are defined as the following:

\begin{align}\label{eq:pseudopotential}
    V_{n,m}= \langle n,n|{}_c\langle M|{}_r\langle m|V|M\rangle_c|m\rangle_r |n,n\rangle
\end{align}

The notation $|n,n\rangle$ represent the two-body state is projected into the $n$-th LL and $|M,m\rangle_r$ is the COM and relative guiding center state of the two body.  For the like-spin Haldane pseudopotential, the two-body state are both particles with the same spin. While for the opposite-spin Haldane pseudopotential, the two-body state is considered to consist of one particle and one hole with different spin since we swap the $\bar2$ and $\bar4$ to get rid of the bar.

In the absence of Landau-level mixing, these bare projected values are identical due to Eq.~\eqref{eq:Coulomb_symmetric}. However, once Landau-level mixing is included, the two pseudopotentials acquire distinct renormalizations, as we will demonstrate in the next section.

\section{Second Order degenerate Perturbation Theory}

In this section we derive closed-form expressions for the second-order 
corrections to the $n-$th LL projected Haldane pseudopotential, as reported in Table I of 
the main text. Our derivation closely follows Ref.~\cite{PhysRevLett.51.605}. 
The second-order perturbation of the 
Coulomb interaction on the $n-$th LL ground-state manifold, as discussed in 
the main text, is given by

\begin{equation}
\mathcal{H}^{\text{eff}} = P_n  V P_n - P_n V P_\perp \frac{1}{ T -E_0} P_\perp V P_n
\equiv \kappa V_n + \kappa^2 \delta V_n +\kappa^2W_n\end{equation}
Here $V_n$ denotes the bare projected Coulomb interaction, $\delta V_n$ the 
second-order effective two-body interaction, and $W_n$ the 
corresponding three-body interaction. For notational convenience, we write the 
Coulomb matrix elements as $V_{1,2';3,4'}^{\sigma\sigma'}$, where $2'=2$ if 
$\sigma=\sigma'$ and $2'=\bar{2}$ if $\sigma'=\bar\sigma$.
We will only focus on the $n=0$ and $n=1$ LL case. 

For the $n=0$ LL, no lower LLs are occupied. 
Hence, the only possible virtual excitations involve promoting 
at least one particle from the $n=0$ LL to higher LLs. 
The resulting effective two-body interaction, denoted $\delta V_0$, 
is referred to as the Bardeen–Cooper–Schrieffer (BCS) correction, 
or equivalently the BCS diagram. 
The explicit formulas for both the two-body and three-body 
effective interactions are presented below. 

\begin{align}
    \delta V_0^{(BCS)} &= -\frac{1}{2}\sum_{1234\sigma\sigma'}\sum_{\substack{n_5+n_6 \geq 1 \\ m_5, m_6}}\frac{V^{\sigma\sigma'}_{1,2';5,6'}V^{\sigma\sigma'}_{5,6';3,4'}}{n_5+n_6}c^{\dagger}_{1\sigma} c^{\dagger}_{2'\sigma'}c_{4'\sigma'}c_{3\sigma} \label{eq:BCS_n0} \\
    W_0 &= -\sum_{\sigma,\sigma',\sigma''}\sum_{1,2',3'',4'',5',6}\sum_{n_7\geq1,m_7} V^{\sigma\sigma'}_{1,2';7,5'}V^{\sigma\sigma''}_{7,3'';6,4''} \frac{1}{n_7}c_{1\sigma}^\dagger c_{2'\sigma'}^\dagger c_{3''\sigma''}^\dagger c_{4''\sigma''} c_{5'\sigma'}c_{6\sigma} \label{eq:three_body_n0}
\end{align}
For the $n=1$ LL, the $n=0$ LL is fully occupied. 
In this case, in addition to exciting at least one particle from the 
$n=1$ LL into higher LLs, one may also excite particles from the 
$n=0$ LL into higher LLs with $n \geq 2$, thereby leaving a hole 
in the $n=0$ LL. Such processes correspond to particle--hole excitations 
and can occur through three distinct channels: the exchange diagram, 
the vertex-correction diagram, and the bubble diagram. Furthermore, 
when two particles from the $n=0$ LL are simultaneously excited 
into the $n=1$ LL, the resulting process corresponds to a hole--hole 
excitation, which is analogous to the BCS 
contribution. For this reason, we present them together. The explicit 
expression for the second-order perturbative correction to the $n=1$ 
ground state is given below.

\begin{align} \label{eq:second_order_correction}
\delta V_1 &=  \sum_{\substack{\alpha = \text{BCS, EX} \\ \text{BB, VC}}} \delta V^{(\alpha)} \\
\delta V^{(\text{BCS})} &= -\frac{1}{2}\sum_{1234\sigma\sigma'}\sum_{\substack{n_5+n_6 \geq 3 \\ m_5, m_6}}\frac{V^{\sigma\sigma'}_{1,2';5,6'}V^{\sigma\sigma'}_{5,6';3,4'}}{n_5+n_6-2}c^{\dagger}_{1\sigma} c^{\dagger}_{2'\sigma'}c_{4'\sigma'}c_{3\sigma} -\frac{1}{2}\sum_{1234 \sigma\sigma'}\sum_{\substack{n_5=n_6=0 \\ m_5, m_6}}\frac{V^{\sigma\sigma'}_{1,2';5,6'}V^{\sigma\sigma'}_{5,6';3,4'}}{2}c^{\dagger}_{1\sigma} c^{\dagger}_{2'\sigma'}c_{4'\sigma'}c_{3\sigma} \label{eq:BCS_diagram}\\
\delta V^{(EX)} &= -\frac{1}{2}\sum_{1234\sigma\sigma'}\sum_{\substack{n_5\geq1,n_6=0 \\ m_5, m_6}}\frac{V^{\sigma\sigma'}_{1,6';5,4'}V^{\sigma\sigma'}_{5,2';3,6'}}{n_5}c^{\dagger}_{1\sigma} c^{\dagger}_{2'\sigma'}c_{4'\sigma'}c_{3\sigma}+(5\leftrightarrow6)\label{eq:EX_diagram}\\
\delta V^{(VC)} &= -\frac{2}{2}\sum_{1234\sigma\sigma'}\sum_{\substack{n_5\geq1,n_6=0 \\ m_5, m_6}}\frac{V^{\sigma\sigma'}_{1,6';3,5'}V^{\sigma\sigma}_{2,5;6,4}}{n_5}c^{\dagger}_{1\sigma} c^{\dagger}_{2'\sigma'}c_{4'\sigma'}c_{3\sigma}+(5\leftrightarrow6)\\
\delta V^{(BB)} &= -\frac{1}{2}\sum_{1234\sigma\sigma'\sigma''}\sum_{\substack{n_5\geq1,n_6=0 \\ m_5, m_6}}\frac{V^{\sigma\sigma''}_{1,5'';3,5''}V^{\sigma''\sigma'}_{6'',2';5'',4'}}{n_5}c^{\dagger}_{1\sigma} c^{\dagger}_{2'\sigma'}c_{4'\sigma'}c_{3\sigma}+(5\leftrightarrow6) \\
W_1 & =-\sum_{\sigma,\sigma',\sigma''}\sum_{1,2',3'',4'',5',6}\sum_{n_7\neq1,m_7} V^{\sigma\sigma'}_{1,2';7,5'}V^{\sigma\sigma''}_{7,3'';6,4''} \frac{1}{n_7-1}c_{1\sigma}^\dagger c_{2'\sigma'}^\dagger c_{3''\sigma''}^\dagger c_{4''\sigma''} c_{5'\sigma'}c_{6\sigma} \label{eq:three_body_n1}
\end{align}

In the following, we focus on the calculation of the effective two-body 
corrections to the Haldane pseudo-potentials. 
Regarding the three-body interaction, we will 
only consider its contribution in the context of the spin-exciton energy, 
which will be addressed separately in Sec.~\ref{sec:three_body_interaction}. 

The basis for the $n$-th Landau-level--projected Haldane pseudopotential is defined in Eq.~\eqref{eq:pseudopotential}. For convenience, we restate it here, after transforming the Landau-level states from the single-particle basis into the center-of-mass and relative basis using Eq.~\eqref{eq:COM basis transformation}
: 
\begin{align}
    |m\rangle_{(n)} &= |n,n\rangle |M\rangle_c|m\rangle_r = \sum_{u=0}^{2n} R^{2n}_{n,u}|(2n-u)M\rangle_c|um\rangle_r
\end{align}
where $R^{L}_{m,n}$ is defined in Eq.~\eqref{eq:R_function} and $|NM\rangle_c|nm\rangle_r$ is the COM and relative state.  The explicit formula for $n=0$  and $n=1$ that we used in the rest of the appendix are given by the following:
\begin{align}
|m\rangle_{(0)} &= |0M\rangle_c|0m\rangle_r \label{eq:n=0_basis}
\\
    |m\rangle_{(1)} &= \frac{1}{\sqrt{2}}\left( |2M\rangle_c|0m\rangle_r-|0M\rangle_c|2m\rangle_r \right) \label{eq:n_1_basis}
\end{align}

\subsection{Correction to Haldane pseudopotential in $n=0$ LL}

We start with the $n=0$ LL. The second order correction to the two-body interaction is given in Eq.~\eqref{eq:BCS_n0}, which we reproduce here,

\begin{align} \label{eq:BCS_def}
  \delta V_0^{(BCS)}=  -\frac{1}{2}\sum_{1234\sigma\sigma'}\sum_{\substack{n_5+n_6 \geq 1 \\ m_5, m_6}}\frac{V^{\sigma\sigma'}_{1,'2;5,6'}V^{\sigma\sigma'}_{5,6';3,4'}}{n_5+n_6}c^{\dagger}_{1\sigma} c^{\dagger}_{2'\sigma'}c_{4'\sigma'}c_{3\sigma}
\end{align}


\paragraph{The like-spin Haldane Pseudopotential:} The detail calculation steps for the BCS diagram is as follows. First, we set $\sigma=\sigma'=\uparrow$ in Eq.~\ref{eq:BCS_def} and evaluate the the matrix element in the basis  Eq.~\ref{eq:n=0_basis} to arrive at the following:


\begin{align}
    \delta V^{\uparrow\uparrow,BCS}_{n=0,m} &={}_{(0)}\langle m|\delta V^{\uparrow\uparrow,BCS}|m\rangle_{(0)} =- \sum_{\substack{n_5+n_6\geq1 \\ m_5, m_6}}{}_{(0)}\langle m|V^{\uparrow\uparrow}|5,6\rangle \frac{1}{n_5+n_6}\langle 5,6|V^{\uparrow\uparrow}|m\rangle_{(0)} \\
    & =-\sum_{n_5+n_6 \geq 1}\left(\sum_{m_5,m_6}{}_c\langle 0M|{}_r\langle 0m|V^{\uparrow\uparrow}|n_5m_5,n_6m_6\rangle \frac{1}{n_5+n_6}\langle n_5m_5,n_6m_6|V^{\uparrow\uparrow}|0M\rangle_c|0m\rangle_r\right)
\end{align}
Next, we use the completeness relation for angular momentum states $\sum_{m_1,m_2}|m_1,m_2\rangle \langle m_1,m_2| = \sum_{J,j}|J\rangle_c|j\rangle_r {}_c\langle J|{}_r\langle j|$ where $|J\rangle_c|j\rangle_r$ is the COM--relative guiding center state and change the single particle Landau level basis into COM -- relative LL basis using Eq.~\eqref{eq:COM basis transformation}. 
This simplify the expression because Coulomb potential is independent of the COM state. As a result, we arrive at the following expression:
\begin{align}
  \delta V^{\uparrow\uparrow,BCS}_{n=0,m} & =- \sum_{n_5+ n_6 \geq 1 }
\sum_{n'=0}^{n_5+n_6}|{}_r\langle 0m|V^{\uparrow\uparrow}|n'(m+n')\rangle_r|^2(R^{n_5+n_6}_{n',n_6})^2\frac{1}{n_5+n_6}\delta_{n_5+n_6-n',0} \\
& =- \sum_{n_5+ n_6 \geq 1 }
|{}_r\langle 0m|V^{\uparrow\uparrow}|(n_5+n_6)(m+n_5+n_6)\rangle_r|^2\frac{(R^{n_5+n_6}_{n_5+n_6,n_6})^2}{n_5+n_6} \label{eq:like_n0}
\end{align}

Here each $(n_5,n_6)$ pair correspond to two virtual particles being excited out of the 0LL. Their contributions to the $\delta V$ is shown in  Fig.~\ref{fig:n0_virtual_state} (a), (c), and (e). They will be used later to contrast the opposite-spin Haldane pseudopotential later.

To proceed further, we perform a change of variables from individual Landau level index $n_5,n_6$ to the total Landau-level index $\alpha$,
\begin{align}
    n_5+n_6 = \alpha \,\ , \,\
    n_6 = \beta.
\end{align}
The new variables are integer values and they satisfy the constrain $\alpha\geq1$ and $\alpha\geq\beta\geq0$ .
After changing the variables, the only dependence on the $\beta$ is in $(R^{n_5+n_6}_{n_5+n_6,n_6})^2 = (R^{\alpha}_{\alpha,\beta})^2$ and summing over the $\beta$ from $0$ to $\alpha$ will give one. Therefore, the expression is simplified into the following form\cite{PhysRevB.87.245425}:

\begin{align}
    \delta V^{\uparrow\uparrow,BCS}_{n=0,m}=- \sum_{\alpha \geq 1 }
|{}_r\langle 0m|V^{\uparrow\uparrow}|\alpha(m+\alpha)\rangle_r|^2\frac{1}{\alpha} = - \frac{(V_{n=0,m}^{\uparrow\uparrow})^2}{4(m+1)}{}_4F_3\left(\substack{1,\,1,\,\frac{3}{2},\,\frac{3}{2} \\ 2,\,2,\,m+2};1 \right)
\end{align}
where $V^{\uparrow\uparrow}_{n=0,m}$ denotes the bare projected Haldane pseudopotential at $n=0$, and ${}_4F_3$ is the generalized hypergeometric function. We evaluate this expression numerically and arrived at the result reported in the maintex.

\paragraph{The opposite-spin Haldane pseudo-potential:} Setting $\sigma=\sigma'=\uparrow$ in Eq.~\ref{eq:BCS_def} leads to the following equation:

\begin{align}
    \delta V^{\uparrow\downarrow,BCS}_{1,\bar2;3,\bar 4}  &= - \sum_{\substack{n_5+n_6 \geq 1 \\ m_5, m_6}} \frac{V^{\uparrow\downarrow}_{1,\bar2;5,\bar6}V^{\uparrow\downarrow}_{5,\bar6;3,\bar4}}{n_5+n_6} = -\sum_{\substack{n_5+n_6 \geq 1 \\ m_5, m_6}} \frac{V^{\uparrow\downarrow}_{1,6;5,2}V^{\uparrow\downarrow}_{5,4;3,6}}{n_5+n_6} = \delta V^{\uparrow\downarrow,BCS}_{1,4;3,2} 
\end{align}
In the above, we used time-reversal property of the LL wavefunctions to simplify the matrix elements, i.e. Eq.~\eqref{eq:Coulomb_symmetric}. When all LL wavefunctions share the same chirality, the same steps as in the like-spin Haldane pseudopotential case can be followed to transform the above equation into the COM and relative LL basis:
\begin{align}
    \delta V^{\uparrow\downarrow,BCS}_{1,4;3,2}& =  \sum_{m,M} \delta V^{\uparrow\downarrow,BCS}_{n=0,m}\,
    R^{M+m}_{m,m_4} R^{M+m}_{m,m_2}\,
    \delta_{m_1+m_4,M+m}\,\delta_{m_3+m_2,M+m}.
\end{align}

\begin{align}
    \delta V^{\uparrow\downarrow,BCS}_{n=0,m} = -\sum_{m_1,m_2,m_3,m_4} R^{M+m}_{m,m_4} R^{M+m}_{m,m_2}\,
    \delta_{m_1+m_4,m_3+m_2}
    \sum_{\substack{n_5+n_6 \geq 1 \\ m_5, m_6}} \frac{V^{\uparrow\downarrow}_{1,6;5,2}V^{\uparrow\downarrow}_{5,4;3,6}}{n_5+n_6}
\end{align}

The Kronecker delta $\delta_{m_1+m_4,m_3+m_2}$ enforces angular momentum conservation between the incoming states $(m_3,m_4)$ and outgoing states $(m_1,m_2)$. Unlike in the like-spin case, the COM and relative LL basis does not simplify the calculation here, since the virtual states $5$ and $6$ do not appear simultaneously in both the ket and the bra. Therefore, we evaluate the matrix elements in momentum space:

\begin{align}
    \langle n_1m_1,n_2m_2|V|n_3m_3,n_4m_4\rangle = \int \frac{d^2q}{(2\pi)^2} V(q)F_{n_1n_3}(\bar q)F_{n_2n_4}(-\bar q) F_{m_1m_3}(q)F_{m_2m_4}(-q)
\end{align}
Here $F_{m_1m_3}$ are the LL form factor in Eq.~\eqref{eq:form_factor} and $V(q) = 2\pi/|q|$. 
Substituting the above expression with Coulomb matrix elements and using the properties of the form factor\cite{macdonald1994introductionphysicsquantumhall}, we arrive at the following form:
\begin{align}
 \delta V^{\uparrow\downarrow,BCS}_{n=0,m}&=   -\sum_{n_5+n_6\geq1} \int \int \frac{dq_1^2dq_2^2}{(2\pi)^4}V_{q_1}V_{q_2} e^{-|q_1+q_2|^2/2}L_m(|q_1+q_2|^2)  F_{0n_5}(q_1)F_{n_50}(q_2)  \frac{1}{n_5+n_6}  F_{0n_6}(-q_2)F_{n_60}(-q_1) \label{eq:opposite_n0}\\
&= -\sum_{n_5+n_6\geq1} \int \int \frac{q_1q_2dq_1dq_2 d\theta_1 d\theta_2}{(2\pi)^4}V_{q_1}V_{q_2} e^{-\frac{q_1^2+q_2^2+2q_1q_2\cos{(\theta_1-\theta_2)}}{2}}L_m(q_1^2+q_2^2+2q_1q_2\cos{(\theta_1-\theta_2)})\notag \\
&\times \frac{(-1)^{n_5+n_6}}{n_5+n_6}  \frac{1}{n_5!n_6!2^{n_5+n_6}} (q_1q_2)^{n_5+n_6}e^{-\frac{q_1^2+q_2^2}{2}} e^{i(\theta_2-\theta_1)(n_5-n_6)}\\
&= -\sum_{n_5+n_6\geq1} \int \int \frac{dq_1dq_2 d\theta}{(2\pi)}e^{-(q_1^2+q_2^2+q_1q_2\cos{\theta})+i\theta(n_5-n_6)}L_m(q_1^2+q_2^2+2q_1q_2\cos{\theta})\frac{(-1)^{n_5+n_6}}{n_5+n_6}  \frac{(q_1q_2)^{n_5+n_6}}{n_5!n_6!2^{n_5+n_6}} 
\end{align}
where $\theta=\theta_1-\theta_2$. For each $(n_5,n_6)$ pair, we evaluate the resulting three-dimensional integral numerically. The contributions from individual $(n_5,n_6)$ pairs are shown in Fig.~\ref{fig:n0_virtual_state}(b)(d)(f). The series converges rapidly with respect to $n_5$ and $n_6$, and summing over all pairs yields the result reported in the main text.

\paragraph{Self energy correction:} We calculate the self energy and its correction since later we will use it in the exciton calculation. The self energy and its correction at a given LL can be calculated from the Haldane pseudo-potentials using the following formula:
\begin{align}
    \Sigma_{nF} &= \sum_{m}\langle nm,nm'|V^{\uparrow\uparrow}|nm',nm\rangle =\int \frac{d^2q}{(2\pi)^2}V(q)|F_{nn}(q)|^2= \sum_{m=0}^{\infty}2(-1)^m V^{\uparrow\uparrow}_{n,m} \label{eq:self_energy} \\
     \delta\Sigma_{nF}^{(\alpha)} &=\sum_{m} 2(-1)^m \delta V_{n,m}^{\uparrow\uparrow,\alpha} 
\end{align}
where $\alpha={BCS,VC,EX,BB}$ are the 4 diagrams.
At the lowest $n=0$ LL, $\Sigma_{0F} = \sqrt{\frac{\pi}{2}}$ and  $\Sigma_{0F}^{(BCS)} \approx -0.6410$. Together with the like-spin and opposite-spin Haldane pseudopotentials and their corresponding corrections, the results are summarized in the table below.

\begin{table*}[h]
    \centering
    \renewcommand{\arraystretch}{1.2}
    \begin{tabular}{|c|cc|cc|cc|cc|cc|c|c|}
        \hline
        Haldane Pseudopotentials & \multicolumn{2}{c|}{$m=0$} & \multicolumn{2}{c|}{$m=1$} & \multicolumn{2}{c|}{$m=2$} & \multicolumn{2}{c|}{$m=3$} & \multicolumn{2}{c|}{$m=4$} & \multicolumn{2}{c|}{Fock Self Energy $\Sigma_F^{(\alpha)}$} \\
        \cline{2-13}
        & $\uparrow\uparrow$ & $\uparrow\downarrow$ & $\uparrow\uparrow$ & $\uparrow\downarrow$
        & $\uparrow\uparrow$ & $\uparrow\downarrow$ & $\uparrow\uparrow$ & $\uparrow\downarrow$
        & $\uparrow\uparrow$ & $\uparrow\downarrow$ & Bare Value &  Correction     \\
        \hline
        \multicolumn{13}{|c|}{\textbf{Lowest Landau level (\(n=0\))}} \\
        \hline
        $V_{0,m}$                & \multicolumn{2}{c|}{0.8862} & \multicolumn{2}{c|}{0.4431}
                                 & \multicolumn{2}{c|}{0.3323} & \multicolumn{2}{c|}{0.2769}
                                 & \multicolumn{2}{c|}{0.2423} & $\sqrt{\frac{\pi}{2}}$ & -0.6410 \\
        $\delta V^{\mathrm{BCS}}_{0,m}$
                                 & -0.3457 & -0.2837 & -0.0328 & -0.0997
                                 & -0.0112 &  0.0036 & -0.0055 & -0.0118
                                 & -0.0033 & -0.0023 &   &  \\
        \hline
    \end{tabular}
    \caption{Spin-resolved Haldane pseudopotentials in the lowest (\(n=0\)). Each \(m\)-sector has two columns corresponding to like-spin (\(\uparrow\uparrow\)) and opposite-spin (\(\uparrow\downarrow\)) Coulomb interactions. The first row gives the bare projected interaction, followed by second-order corrections from the indicated diagrammatic processes. In the last two column, we show the Fock self energy.}
    \label{tab:haldane_pseudopotentials_n0}
\end{table*}

\subsection{Comparison between Like-Spin and Opposite-Spin Haldane Pseudopotentials in the $n=0$} 
\begin{figure}[h]
    \centering
    \includegraphics[width=1\linewidth]{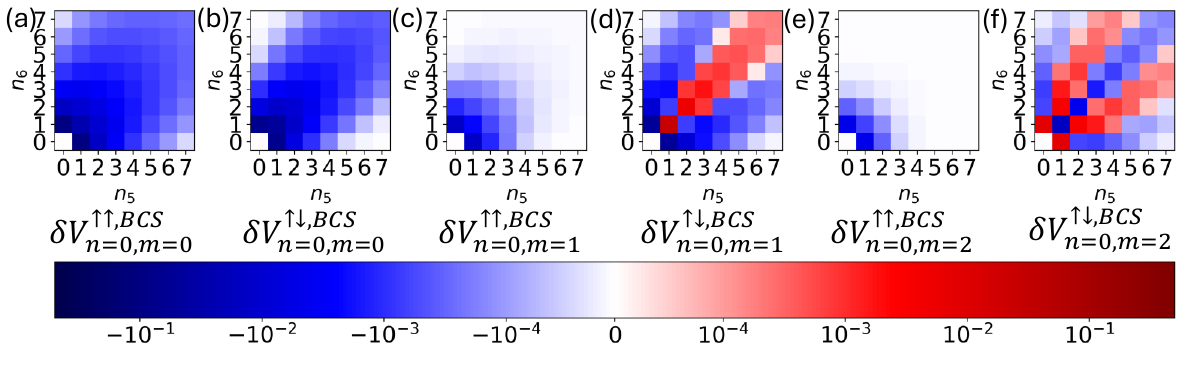}
    \caption{Each panel shows the contribution of virtual states to the Haldane 
pseudo-potential in the $n=0$ Landau level (LL). Panels~(a)--(b) correspond 
to the $m=0$ components of the like-spin and opposite-spin pseudo-potentials, 
respectively, while panels~(c)--(d) correspond to $m=1$, and panels~(e)--(f) 
to $m=2$. The horizontal and vertical axes denote the Landau-level indices 
of the intermediate virtual states, labeled $n_5$ and $n_6$ in the second-order 
expressions. The sum of all contributions within each colormap yields the corresponding 
 correction reported in Table~I of the main text. 
}
    \label{fig:n0_virtual_state}
\end{figure}
Fig.~\ref{fig:n0_virtual_state} shows the contribution of each virtual state to the Haldane pseudopotentials. The axes correspond to the Landau-level indices of the virtual states, $n_5$ and $n_6$, as defined in Eqs.\eqref{eq:like_n0} and \eqref{eq:opposite_n0}.

Fig.~\ref{fig:n0_virtual_state}a) and b) compares the second order correction between the effective $m=0$ like-spin and opposite-spin Haldane pseudopotential.
The phase space are identical between the effective like-spin and opposite-spin as both have $n_5+n_6\geq1$, but their matrix elements are different, as derived above. The differences in their matrix elements lead to stronger screening in the like-spin than the opposite spin, as shown in the colormap. As shown in Tab.~\ref{tab:haldane_pseudopotentials_n0}, the $\delta V_{n=0,m=0}^{\uparrow\uparrow,BCS} = -0.3467$ and $\delta V_{n=0,m=0}^{\uparrow\downarrow,BCS} = -0.2837$. 

In the $m=1$ case, Fig.~\ref{fig:n0_virtual_state}c) and d), the individual contribution in the like-spin  channel decrease rapidly compared to those at $m=0$, whereas the values in 
the opposite-spin channel remain relatively large. Notably, the opposite-spin 
channel exhibits certain positive contributions, which are absent in the 
like-spin case. 

In the like-spin sector, the virtual states share the 
same chirality, and the BCS diagram always reduces the interaction strength, 
$\delta V^{BCS,\uparrow\uparrow}<0$.

By contrast, when the virtual states have opposite chirality, it is no longer guaranteed that the BCS 
channel will reduce the interaction strength, again due to the matrix element effects. 
Although positive contributions are 
indeed present shown in Fig.~\ref{fig:n0_virtual_state}(d), they diminish as the virtual states move farther away from the projected states.

The virtual state next to the projected state still provides 
a negative contribution and dominates the screening effect. Since the 
overall magnitude of individual contribution in the opposite-spin channel is 
much larger than that in the like-spin channel, this ultimately results in 
stronger screening for the opposite-spin interaction with $\delta V_{n=0,m=1}^{\uparrow\uparrow,BCS} = -0.0328$ and $\delta V_{n=0,m=1}^{\uparrow\downarrow,BCS} = -0.0997$.

For $m=2$, the individual contributions to $\delta V^{\uparrow\uparrow}_{n=0,m=2}$ are further suppressed compared to both $\delta V^{\uparrow\uparrow}_{n=0,m=1}$ and $\delta V^{\uparrow\uparrow}_{n=0,m=0}$. In contrast, the opposite-spin channel again retains sizable contributions. The dominant effect arises from virtual states with the smallest energy denominators, namely $(n_5,n_6)=(1,0)$ and $(0,1)$. These channels yield positive contributions, leading to an overall positive correction to $\delta V^{\uparrow\downarrow}_{n=0,m=2}$. Such behavior is absent in conventional quantum Hall systems~\cite{PhysRevB.87.245425}, where all Haldane pseudopotentials are reduced by screening.

A subtle feature emerges in the comparison between like-spin and opposite-spin Haldane pseudopotentials in the $n=0$ LL: an even–odd effect. Even-$m$ components are more strongly screened in the like-spin channel, while odd-$m$ components are more strongly screened in the opposite-spin channel. We examined the first few Haldane pseudopotentials but found no clear physical mechanism underlying this behavior. It remains an open question deserving further study.

\subsection{Correction to The Like-spin Haldane pseudopotential in $n=1$ LL}
In this section, we calculate the correction to the like-spin Haldane pseudo-potential $\delta V_{n=1,m}^{\uparrow\uparrow}$ arising from the effective two-body interaction. We will talk about the four distinct diagrams separately in the following.

\paragraph{BCS Diagram:}

The expression for the BCS diagrams in the like-spin channel, given in Eq.~\eqref{eq:BCS_diagram} for $\sigma=\sigma'=\uparrow$, can be transfomed into the COM and relative angular momentum basis in $n=1$ using Eq.~\eqref{eq:n_1_basis}. In this representation, the correction takes the form:

\begin{align}
    \delta V^{\uparrow\uparrow,BCS}_{m,n=1} &={}_{(1)}\langle m|\delta V^{\uparrow\uparrow,BCS}|m\rangle_{(1)} \\
    &=- \sum_{\substack{n_5, n_6 \\ m_5, m_6}}{}_{(1)}\langle m|V^{\uparrow\uparrow}|5,6\rangle \frac{\theta(n_5,n_6)}{n_5+n_6-2}\langle 5,6|V^{\uparrow\uparrow}|m\rangle_{(1)} - \frac{1}{2} \sum_{m_5, m_6}{}_{(1)}\langle m|V^{\uparrow\uparrow}|5,6\rangle \langle 5,6|V^{\uparrow\uparrow}|m\rangle_{(1)}\big|_{n_5=n_6=0} \label{eq:pp and hh}
\end{align}
The function $\theta(n_5,n_6)$ restrict the summation domain to $n_5\geq1$, $n_6\geq1$ and $n_5+n_6\geq3$. This means one of the virtual LL must be outside of the $n=1$ manifold.  
The first term in Eq.~\eqref{eq:pp and hh} corresponds to the virtual particle-particle state contribution, which we denote as \( \delta V_{m,n=1}^{\uparrow\uparrow,\mathrm{BCS}}(pp) \); the second term corresponds to the virtual hole-hole state contribution, labeled \( \delta V_{m,n=1}^{\uparrow\uparrow,\mathrm{BCS}}(hh) \). We will evaluate them separately.

We begin by considering the particle--particle contribution  
\( \delta V_{m,n=1}^{\uparrow\uparrow,\mathrm{BCS}}(pp) \). 
Since the Haldane pseudopotential basis at $n=1$ is expressed as a superposition 
of two COM and relative states, the resulting expression naturally 
separates into two square terms and two identical cross terms. 
Accordingly, we decompose the correction as
\begin{equation}
\delta V^{\uparrow\uparrow,\mathrm{BCS}}_{m,n=1}(pp) 
= \delta V^{\mathrm{BCS}}_1 + 2\delta V^{\mathrm{BCS}}_2 + \delta V^{\mathrm{BCS}}_3 ,
\end{equation}
where \( \delta V^{\mathrm{BCS}}_1 \) and \( \delta V^{\mathrm{BCS}}_3 \) represent the square terms, 
and \( \delta V^{\mathrm{BCS}}_2 \) corresponds to the cross term. 
Each contribution is evaluated individually.

For \( \delta V^{\mathrm{BCS}}_1 \), the expression is given by:
\begin{align}
    \delta V^{BCS}_1 &= \frac{1}{2}\sum_{n_5,n_6 \geq 1}\sum_{m_5,m_6}{}_c\langle M,2|{}_r\langle m,0|V^{\uparrow\uparrow}|n_5m_5,n_6m_6\rangle \frac{\theta(n_5,n_6)}{2-n_5-n_6}\langle n_5m_5,n_6m_6|V^{\uparrow\uparrow}|M,2\rangle_c|m,0\rangle_r\\
    & = \frac{1}{2}\sum_{n_5,n_6 \geq 1}\sum_{J,j}\sum_{n'',n'=0}^{n_5+n_6}{}_c\langle M,2|{}_r\langle m,0|V^{\uparrow\uparrow}|J,(n_5+n_6-n')\rangle_c|j,n'\rangle_rR^{n_5+n_6}_{n',n_6} \notag \\
    &\times \frac{\theta'(n_5,n_6)}{2-n_5-n_6} R^{n_5+n_6}_{n'',n_6}{}_c\langle J,(n_5+n_6-n'')|{}_r\langle j,n''|V^{\uparrow\uparrow}|M,2\rangle_c|m,0\rangle_r\\
    & = \frac{1}{2}\sum_{\substack{n_5, n_6 \geq 1 \\ n_5 + n_6 \geq 3}}
\sum_{n'=0}^{n_5+n_6}|{}_r\langle m,0|V^{\uparrow\uparrow}|n'+m,n'\rangle_r|^2(R^{n_5+n_6}_{n',n_6})^2\frac{1}{2-n_5-n_6}\delta_{n_5+n_6-n',2}
\end{align}
Analogous to the BCS correction in the $n=0$ like-spin Haldane pseudopotential, we make use of  the completeness relation for guiding-center states,  
$
\sum_{m_1,m_2}|m_1,m_2\rangle\langle m_1,m_2| 
= \sum_{J,j}|J\rangle_c|j\rangle_r\,{}_c\langle J|\,{}_r\langle j| ,
$
together with the transformation of the single-particle Landau-level basis into the 
COM and relative basis via Eq.~\eqref{eq:COM basis transformation} in deriving the second line from the first.

Next, we change variables from the single-particle indices \((n_5,n_6)\) to 
the total Landau-level index \(\alpha=n_5+n_6\) (with \(\alpha\geq 3\)) and 
\(n_6=\beta\) (with \(1\leq\beta\leq\alpha-1\)). 
Using the orthogonality property of the coefficients $R^L_{m,n}$, namely 
\(\sum_{n=0}^{L}R^{L}_{m',n}R^{L}_{m,n}=\delta_{m m'}\), 
we arrive at a simplified expression for \(\delta V^{\mathrm{BCS}}_1\).
\begin{align}
     \delta V^{BCS}_1 
     &=\frac{1}{2}\sum_{\alpha \geq 3}\sum_{\beta \geq 1}^{\alpha-1}|{}_r\langle m,0|V^{\uparrow\uparrow}|\alpha+m-2,\alpha-2\rangle_r|^2\frac{(R^{\alpha}_{\alpha-2,\beta})^2}{2-\alpha}\\
     & = \frac{1}{2}\sum_{\alpha \geq 3}\big|{}_r\langle m,0|V^{\uparrow\uparrow}|\alpha+m-2,\alpha-2\rangle_r\big|^2\frac{1}{2-\alpha} \left(1-(R^{\alpha}_{\alpha-2,\alpha})^2-(R^{\alpha}_{\alpha-2,0})^2 \right)\\
     & = -\frac{1}{2}\sum_{\alpha \geq 1}\big|{}_r\langle m,0|V^{\uparrow\uparrow}|\alpha+m,\alpha\rangle_r\big|^2\frac{1}{\alpha}+ \frac{1}{2}\sum_{\alpha \geq 1}\big|{}_r\langle m,0|V^{\uparrow\uparrow}|\alpha+m,\alpha\rangle_r\big|^2\frac{1}{\alpha} \frac{(\alpha+2)(\alpha+1)}{2^{\alpha+2}}
\end{align}

The cross term, \(\delta V^{\mathrm{BCS}}_2\), is computed in complete analogy, 
again invoking the COM--relative transformation:

\begin{align}
     \delta V^{BCS}_2 &=- \frac{1}{2}\sum_{n_5,n_6 \geq 1}\sum_{m_5,m_6}{}_c\langle M,2|{}_r\langle m,0|V^{\uparrow\uparrow}|n_5m_5,n_6m_6\rangle \frac{\theta(n_5,n_6)}{2-n_5-n_6}\langle n_5m_5,n_6m_6|V^{\uparrow\uparrow}|M,0\rangle_c|m,2\rangle_r\\
     & = -\frac{1}{2}\sum_{n_5,n_6 \geq 1}\sum_{J,j}\sum_{n'',n'}^{n_5+n_6}{}_c\langle M,2|{}_r\langle m,0|V^{\uparrow\uparrow}|J,(n_5+n_6-n')\rangle_c|j,n'\rangle_rR^{n_5+n_6}_{n',n_6} \notag \\
    &\times \frac{\theta(n_5,n_6)}{2-n_5-n_6} R^{n_5+n_6}_{n'',n_6}{}_c\langle J,(n_5+n_6-n'')|{}_r\langle j,n''|V^{\uparrow\uparrow}|M,0\rangle_c|m,2\rangle_r\\
    & = - \frac{1}{2}\sum_{n_5,n_6 \geq 1}\sum_{n'}^{n_5+n_6}{}_r\langle m,0|V^{\uparrow\uparrow}|m+n',n'\rangle_rR^{n_5+n_6}_{n',n_6}\delta_{n_5+n_6-n',2} \frac{\theta(n_5,n_6)}{2-n_5-n_6} R^{n_5+n_6}_{n'+2,n_6}{}_r\langle m+n',n'+2|V^{\uparrow\uparrow}|m,2\rangle_r
\end{align}
Using orthogonality of the 
$R^L_{m,n}$ coefficients and a similar change of variables yields the compact expression in the following:

\begin{align}
   \delta V^{BCS}_2 &=\frac{1}{2}\sum_{\alpha \geq 3}\sum_{\beta\geq1}^{\alpha-1}{}_r\langle m,0|V^{\uparrow\uparrow}|m+\alpha-2,\alpha-2\rangle_rR^{\alpha}_{\alpha-2,\beta} \frac{1}{2-\alpha} R^{\alpha}_{\alpha,\beta}{}_r\langle m+\alpha-2,\alpha|V^{\uparrow\uparrow}|m,2\rangle_r \\
    & = -\frac{1}{2}\sum_{\alpha \geq 3}{}_r\langle m,0|V^{\uparrow\uparrow}|m+\alpha-2,\alpha-2\rangle_r \frac{1}{2-\alpha} {}_r\langle m+\alpha-2,\alpha|V^{\uparrow\uparrow}|m,2\rangle_r\left( R^{\alpha}_{\alpha-2,0}R^{\alpha}_{\alpha,0}+ R^{\alpha}_{\alpha-2,\alpha}R^{\alpha}_{\alpha,\alpha} \right)\\
    & = -\frac{1}{2}\sum_{\alpha \geq 3}{}_r\langle m,0|V^{\uparrow\uparrow}|m+\alpha-2,\alpha-2\rangle_r \frac{1}{2-\alpha} {}_r\langle m+\alpha-2,\alpha|V^{\uparrow\uparrow}|m,2\rangle_r \frac{1}{2^{\alpha-1}}\sqrt{\frac{\alpha(\alpha-1)}{2}}\\
    & = \frac{1}{2}\sum_{\alpha \geq 1}{}_r\langle m,0|V^{\uparrow\uparrow}|m+\alpha,\alpha\rangle_r \frac{1}{\alpha} {}_r\langle m+\alpha,\alpha+2|V^{\uparrow\uparrow}|m,2\rangle_r \frac{1}{2^{\alpha+1}}\sqrt{\frac{(\alpha+2)(\alpha+1)}{2}}
\end{align}

As for $\delta V^{BCS}_{3}$, applying the same procedure yields the following form:
\begin{align}
    \delta V^{BCS}_{3} = -\frac{1}{2}\sum_{\alpha\geq1}\big|{}_r\langle m,2|V^{\uparrow\uparrow}|m+\alpha,\alpha+2\rangle_r\big|^2 \frac{1}{\alpha}+\frac{1}{2}\sum_{\alpha\geq1}\big|{}_r\langle m,2|V^{\uparrow\uparrow}|m+\alpha,\alpha+2\rangle_r\big|^2 \frac{1}{\alpha}\frac{1}{2^{\alpha+1}}
\end{align}

Collecting all three parts, we obtain the $\delta V^{\uparrow\uparrow,BCS}_{m,n=1}(pp)$:

\begin{align}\label{eq:BCS n=1 like-spin}
    \delta V^{\uparrow\uparrow,BCS}_{n=1,m}(pp) &= \delta V^{\mathrm{BCS}}_1 + 2\delta V^{\mathrm{BCS}}_2 + \delta V^{\mathrm{BCS}}_3 \\
    &=-\frac{1}{2}\sum_{\alpha \geq 1}\big|{}_r\langle m,0|V^{\uparrow\uparrow}|\alpha+m,\alpha\rangle_r\big|^2\frac{1}{\alpha}+ \frac{1}{2}\sum_{\alpha \geq 1}\big|{}_r\langle m,0|V^{\uparrow\uparrow}|\alpha+m,\alpha\rangle_r\big|^2\frac{1}{\alpha} \frac{(\alpha+2)(\alpha+1)}{2^{\alpha+2}}\\
    &-\sum_{\alpha \geq 1}{}_r\langle m,0|V^{\uparrow\uparrow}|m+\alpha,\alpha\rangle_r \frac{1}{\alpha} {}_r\langle m+\alpha,\alpha+2|V^{\uparrow\uparrow}|m,2\rangle_r \frac{1}{2^{\alpha+1}}\sqrt{\frac{(\alpha+2)(\alpha+1)}{2}}\\
    &-\frac{1}{2}\sum_{\alpha\geq1}\big|{}_r\langle m,2|V^{\uparrow\uparrow}|m+\alpha,\alpha+2\rangle_r\big|^2 \frac{1}{\alpha}+\frac{1}{2}\sum_{\alpha\geq1}\big|{}_r\langle m,2|V^{\uparrow\uparrow}|m+\alpha,\alpha+2\rangle_r\big|^2 \frac{1}{\alpha}\frac{1}{2^{\alpha+1}}
\end{align}

Finally, for the hole-hole contribution $\delta V^{\uparrow\uparrow,BCS}_{m,n=1}(hh)$, its expression is given by:
\begin{align}
   \delta V^{\uparrow\uparrow,BCS}_{m,n=1}(hh)=- \frac{1}{2} \sum_{m_5, m_6}{}_{(1)} \langle m|V^{\uparrow\uparrow}|0m_5,0m_6\rangle \langle 0m_5,0m_6|V^{\uparrow\uparrow}|m\rangle_{(1)}
\end{align}
This also have in general four terms because of the superposition of the basis state.
But since $n_5=n_6=0$,the only possible COM and relative state is also $|0\rangle_c|0\rangle_r$ and the Coulomb interaction conserves the COM index, only the second square term with $|N=0\rangle_c|n=2\rangle_r$ will be nonzero. Following the same steps as $V^{\uparrow\uparrow,BCS}_{m,n=1}(pp)$, the expression for $\delta V^{\uparrow\uparrow,BCS}_{m,n=1}(hh)$ is given by the following:
\begin{align}
    \delta V^{\uparrow\uparrow,BCS}_{m,n=1(hh)}=- \frac{1}{4} \sum_{J,j}|{}_c\langle 0,M|{}_r\langle 2,m|V^{\uparrow\uparrow}|0J\rangle_c|0j\rangle_r|^2 = -\frac{1}{4}|{}_r\langle 0,m-2|V^{\uparrow\uparrow}|2,m\rangle_r|^2 \label{eq:hh_con}
\end{align}
From the above expression, we see that due to angular momentum conservation, the hole-hole contribution does not contribute to $m=0$ and $m=1$ like-spin Haldane pseudo-potential components. Combining both the particle–particle and hole–hole contributions, we obtain the BCS correction reported in Table I of the main text.

\paragraph{Exchange Diagram:} We evaluate the exchange-diagram correction to the like-spin Haldane pseudopotential, 
denoted by \(\delta V_{n=0,m}^{\uparrow\uparrow,EX}\). 
The  matrix elements of the exchange diagram are given by :

\begin{align}
    \delta V^{\uparrow\uparrow,EX}_{1234} =\langle 1\uparrow,2\uparrow|\delta V^{\uparrow\uparrow,EX}|3\uparrow,4\uparrow\rangle=\sum_{5,6}V^{\uparrow\uparrow}_{16,54}\frac{\pi(n_5,n_6)}{n_5-n_6}V^{\uparrow\uparrow}_{52,36}
     = \sum_{5,6}\langle 1,6|V^{\uparrow\uparrow}|5,4\rangle \frac{\pi(n_5,n_6)}{n_5-n_6}\langle 5,2|V^{\uparrow\uparrow}|3,6\rangle \label{eq:first quantization}
\end{align}
where all external Landau-level indices are fixed at \(n=1\). 
The function \(\pi(n_5,n_6)\) restricts the summation domain to \(n_5 \geq 1\) and \(n_6=0\). 
Unlike in the BCS diagram, transforming the states within the same bra and ket into the COM 
and relative basis does not lead to simplifications, since external and virtual states 
appear mixed within each bra or ket. 
For this reason, it is more convenient to evaluate the exchange diagram directly in momentum 
space, using the form factor defined in Eq.~\eqref{eq:form_factor}. This calculation is closely related to the opposite-spin BCS case. In fact, the Coulomb matrix elements are identical between the like-spin exchange diagram and the opposite-spin BCS diagram due to the chirality properties ofwavefunction ; the only differences lie in the energy denominator and the summation range. We will make use of this correspondence to evaluate the opposite-spin correction in the next section.

The Coulomb interaction in momentum space is given by:
\begin{align}
    V^{\uparrow\uparrow} = \sum_{q} V_{q}e^{iq(r_i-r_j)}, \,\ V_q=\frac{2\pi}{|q|}
\end{align}
where we absorb the factor $1/(2A)$ into the $\sum_q$.
Here, $\mathbf{r}_i$ and $\mathbf{r}_j$ denote distinct position operators.  
Using the Coulomb interaction in momentum space, the matrix elements of the exchange diagram can be written as:

\begin{align}
    \delta V^{\uparrow\uparrow,EX}_{1234}&= \sum_{5,6}\sum_{q_1,q_2} V_{q_1}V_{q_2}\langle 1,6|e^{iq_1(r_i-r_j)}|5,4\rangle \frac{\pi(n_5,n_6)}{n_5-n_6}\langle 5,2|e^{iq_2(r_i-r_j)}|3,6\rangle
\end{align}
The corresponding exchange correction to the Coulomb interaction is given by:

\begin{align}
     \delta V^{\uparrow\uparrow,EX}&=\sum_{q_1,q_2}V_{q_1}V_{q_2} \sum_{n_5n_6}\sum_{m_5m_6}(e^{iq_1 r_i}|n_5m_5\rangle\langle n_5m_5 |e^{iq_2 r_i}) \frac{\pi(n_5,n_6)}{n_5-n_6}( e^{-iq_2 r_j} |n_6m_6\rangle\langle n_6m_6|e^{-iq_1r_j})
\end{align}

The exchange correction to like-spin Haldane pseudopotential is given by $V^{\uparrow\uparrow, \mathrm{EX}}_{m,n=1} = {}_{(1)}\langle m|\, \delta V^{\uparrow\uparrow, \mathrm{EX}}\, |m\rangle_{(1)} $, but
before computing $V^{\uparrow\uparrow, \mathrm{EX}}_{m,n=1}$, we can perform several simplifications.  
First, the summation over $m_5$ and $m_6$ yield a completeness relation in guiding center space.  
Additionally, the position operator $\vec{r}$ can be decomposed into the cyclotron and guiding center coordinates $\vec{r} = \vec{\eta} + \vec{R}$,
where $\vec{\eta}$ is the cyclotron coordinate and $\vec{R}$ is the guiding center coordinate.  
Using this decomposition, the expression can be rewritten as:

\begin{align}
    \delta V^{\uparrow\uparrow,EX}&=\sum_{q_1,q_2}V_{q_1}V_{q_2} \sum_{n_5}\sum_{n_6}(e^{iq_1(\eta_i+R_i)}|n_5\rangle\langle n_5|e^{iq_2(\eta_i+R_i)})  \frac{\pi(n_5,n_6)}{n_5-n_6}( e^{-iq_2(\eta_j+R_j)} |n_6 \rangle\langle n_6|e^{-iq_1(\eta_j+R_j)})
\end{align}

Since $\vec{\eta}$ acts only on the LL quantum states and $\vec{R}$ acts only on the guiding center states, so we can treat the LL and guiding center separately. Let's first project $\delta V^{\uparrow\uparrow,EX}$ into the $n=1$ LL, this yields the following expression:

\begin{align}
    \delta V^{\uparrow\uparrow, EX}_{n=1}
    & = \langle n=1,n=1|\delta V^{\uparrow\uparrow,EX}|n=1,n=1\rangle\\
    & = \sum_{q_1,q_2}V_{q_1}V_{q_2}   \sum_{n_5}\sum_{n_6}(F_{1n_5}(q_1)F_{n_51}(q_2))  \frac{\pi(n_5,n_6)}{n_5-n_6}  (F_{1n_6}(-q_2)F_{n_61}(-q_1) ) e^{i(q_1+q_2)(R_i-R_j)}
\end{align}

The form factor $F_{n'n}$ is defined in Eq.~\eqref{eq:form_factor} and satisfies the relation $F_{nn'}(\vec{q}) = F^*_{n'n}(-\vec{q})$. In the above equation, the operator $e^{i(q_1+q_2)(R_i-R_j)}$ only acts on the relative guiding center states. Taking the expectation value of this operator on the state  $|m\rangle_r$ lead to the following:

\begin{align}
    \delta V^{\uparrow\uparrow, EX}_{n=1,m} &= {}_r\langle m| \delta V^{\uparrow\uparrow, EX}_{n=1} | m\rangle_r \\
    & = \sum_{q_1,q_2}V_{q_1}V_{q_2}   \sum_{n_5}\sum_{n_6}(F_{1n_5}(q_1)F_{n_51}(q_2))  \frac{\pi(n_5,n_6)}{n_5-n_6}  (F_{1n_6}(-q_2)F_{n_61}(-q_1) ) e^{-|q_1+q_2|^2/2} L_m(|q_1+q_2|^2)
\end{align}

In the above expression, we have used the identity
${}_r\langle m| e^{i\vec{q} \cdot (\vec{R}_i - \vec{R}_j)} |m\rangle_r = e^{-|\vec{q}|^2/2} L_m(|\vec{q}|^2)$.
With these ingredients, we arrive at the final integral expression for the exchange diagram contribution to the like-spin Haldane pseudo-potential:
\begin{align}
    \delta V_{m,n=1}^{\uparrow\uparrow,EX} 
    &= \int \int \frac{dq_1^2dq_2^2}{(2\pi)^4}V_{q_1}V_{q_2} e^{-|q_1+q_2|^2/2}L_m(|q_1+q_2|^2) \sum_{n_5,n_6}F_{1n_5}(q_1)F_{n_51}(q_2)  \frac{\pi(n_5,n_6)}{n_5-n_6}  F_{1n_6}(-q_2)F_{n_61}(-q_1)\\
    &=\int \int \frac{dq_1^2dq_2^2}{(2\pi)^4}V_{q_1}V_{q_2} e^{-|q_1+q_2|^2/2}L_m(|q_1+q_2|^2)  F_{10}(-q_2)F_{01}(-q_1) \sum_{n\geq1}\frac{F_{1n}(q_1)F_{n1}(q_2)}{n} \label{eq:ph_ex}
\end{align}

In the above expression, we have used the property of \(\pi(n_5,n_6)\), 
which restricts the summation to \(n_5 \geq 1\) and \(n_6 = 0\). 
In addition, there is an identical contribution obtained by exchanging \(n_5\) and \(n_6\). 
A direct numerical evaluation of this expression yields the exchange correction 
to the like-spin Haldane pseudopotential.

\paragraph{Vertex correction diagram}
The vertex-correction diagram follows the same reasoning as the exchange diagram, 
though with some differences in detail. 
The corresponding matrix elements are given by:

\begin{align}
    \delta V^{\uparrow\uparrow,VC}_{1234} =\sum_{5,6}\sum_{q_1,q_2}V_{q_1}V_{q_2} \langle 1|e^{iq_2r_i}|5\rangle\langle5|e^{-iq_1r_i}|6\rangle \langle6|e^{-iq_2r_i}|3\rangle \langle 2|e^{iq_1r_j}|4\rangle \frac{\pi(n_5,n_6)}{n_5-n_6}
\end{align}
with the vertex correction to the Coulomb interaction gievn by:
\begin{align}
    \delta V^{\uparrow\uparrow,VC} =\sum_{5,6}\sum_{q_1,q_2}V_{q_1}V_{q_2}e^{iq_2r_i}|5\rangle\langle5|e^{-iq_1r_i}|6\rangle \langle6|e^{-iq_2r_i}   e^{iq_1r_j}\rangle \frac{\pi(n_5,n_6)}{n_5-n_6}
\end{align}

The summation over \(m_5\) and \(m_6\) is straightforward and yields the completeness relation. 
Next, we decompose the position operator as \(r=\eta+R\), where \(\eta\) is the cyclotron coordinate 
and \(R\) is the guiding-center coordinate, as in the exchange-diagram calculation. 
Projecting the vertex correction of the Coulomb interaction onto the first Landau level then leads to:
\begin{align}
        \delta V_{n=1}^{\uparrow\uparrow,VC}&= \langle n=1,n=1|\delta V^{\uparrow\uparrow,VC}|n=1,n=1\rangle \\
        &= \sum_{q_1,q_2}\sum_{n_5,n_6} V_{q_1}V_{q_2}F_{1n_5}(q_2)F_{n_5n_6}(-q_1)F_{n_61}(-q_2)F_{11}(q_1)\frac{\pi(n_5,n_6)}{n_5-n_6}e^{iq_2R_i} e^{-iq_1R_i}e^{-iq_2R_i}e^{iq_1R_j}
\end{align}
Using the commutation relations of the guiding-center coordinates $e^{iq_2R_i} e^{-iq_1R_i}e^{-iq_2R_i}e^{iq_1R_j} = e^{i \hat{z}\cdot (q_1\times q_2)} e^{-iq_1(R_i-R_j)}$
, the above expression simplifies to

\begin{align}
     \delta V_{n=1}^{\uparrow\uparrow,VC} = \sum_{q_1,q_2}\sum_{n_5,n_6} V_{q_1}V_{q_2}F_{1n_5}(q_2)F_{n_5n_6}(-q_1)F_{n_61}(-q_2)F_{11}(q_1)\frac{\pi(n_5,n_6)}{n_5-n_6}e^{i \hat{z}\cdot (q_1\times q_2)} e^{-iq_1(R_i-R_j)}
\end{align}

Thus, \(\delta V_{n=1}^{\uparrow\uparrow,\mathrm{VC}}\) depends only on the relative guiding-center coordinates. 
Consequently, we can evaluate the vertex-correction diagram to the Haldane pseudopotential 
by sandwiching it with the relative basis \(|m\rangle_r\). 
The vertex-correction contribution, denoted by \(\delta V^{\uparrow\uparrow,\mathrm{VC}}_{m,n=1}\), is given by:

\begin{align}
    \delta V^{\uparrow\uparrow,VC}_{m,n=1} &= {}_{r}\langle m|\delta V^{\uparrow\uparrow,VC}_{n=1}|m\rangle_{r}\\
    & = \sum_{q_1,q_2}\sum_{n_5,n_6} V_{q_1}V_{q_2}F_{1n_5}(q_2)F_{n_5n_6}(-q_1)F_{n_61}(-q_2)F_{11}(q_1)\frac{\pi(n_5,n_6)}{n_5-n_6} e^{-\frac{|q_1|^2}{2}}L_m(|q_1|^2)e^{i\hat z \cdot(q_1\times q_2)}\\
    & = \int\int \frac{d^2q_1d^2q_2}{(2\pi)^4}V_{q_1}V_{q_2}(\sum_{n_5\geq1}\frac{F_{1n_5}(q_2)F_{n_50}(-q_1)}{n_5})F_{01}(-q_2)F_{11}(q_1)e^{-\frac{|q_1|^2}{2}}L_m(|q_1|^2)e^{i\hat z \cdot(q_1\times q_2)} \label{eq:ph_vc_like}
\end{align}

The above integral is evaluated numerically. 
Since there are four identical contributions, their sum yields the vertex correction 
reported in Table~I of the main text.

\paragraph{Bubble diagram}

Finally, we consider the bubble diagram contribution. Following the same procedure outlined in the previous calculations, the bubble correction to the Coulomb interaction, denoted by $\delta V^{\uparrow\uparrow,BB}$, is given by:

\begin{align}
   \delta V^{\uparrow\uparrow,BB}& = \sum_{q_1,q_2}\sum_{5,6} V_{q_1}V_{q_2}e^{iq_1r_1}e^{-iq_2r_2} \langle 5| e^{-iq_1r}|6\rangle \langle 6| e^{iq_2r}|5\rangle \frac{\pi(n_5,n_6)}{n_5-n_6}
\end{align}

Unlike the calculation before, the summation over $m_5$ and $m_6$ yields a delta function $\delta_{q_1 q_2}$\cite{macdonald1994introductionphysicsquantumhall}.
As a result, the above expression can be further simplified to the following form:

\begin{align}
    \delta V^{\uparrow\uparrow,BB} = \sum_{q} V_{q}^2 \sum_{n_5\geq 1} \frac{|F_{n_50}(q)|^2}{n_5}e^{iq(r_1-r_2)}
\end{align}
The bubble diagram correction to the Haldane pseudopotential takes the following simplified form:
\begin{align}
    \delta V^{\uparrow\uparrow,BB}_{m,n=1} &=\sum_{q} V_{q}^2 \sum_{n_5\geq 1} \frac{|F_{n_50}(q)|^2}{n_5}{}_{(1)}\langle m|e^{iq(r_1-r_2)}|m\rangle_{(1)}\\
    & = \sum_{q} V_{q}^2 \sum_{n_5\geq 1} \frac{|F_{n_50}(q)|^2}{n_5} (F_{11}(q))^2 L_m(|q|)e^{-\frac{|q|^2}{2}}  \label{eq:ph_bb_like}    \\
    & =\sum_{q} V_{q}^2 e^{-\frac{|q|^2}{2}} \left( \mathrm{Ei}(\frac{q^2}{2}) +\mathrm{ln}(\frac{q^2}{2})-\gamma \right) (F_{11}(q))^2 L_m(|q|)e^{-\frac{|q|^2}{2}}
\end{align}
Here, $\mathrm{Ei}(x)$ denotes the exponential integral function and $\gamma$ is the Euler--Mascheroni constant. Note that there are four additional identical contribution arising from the closed fermi loop and the exchange of $n_5$ and $n_6$. Evaluating the expression numerically gives the corresponding correction reported in the maintext.

\paragraph{Self energy}
Here,  we calculate the total exchange energy $\Sigma_{n=1,F}$ to second order in $\kappa$ for the $n=1$ LL.
Applying the same formulas in Eq.~\ref{eq:self_energy}, we have:

\begin{align}
    \Sigma_{1F} = \sum_{m} 2(-1)^m V^{\uparrow\uparrow}_{n=1,m} = \int d^2q V(q)|F_{11}(q)|^2=\sqrt{\frac{\pi}{2}}\frac{3}{4}
\end{align}
where $F_{11}(q)=(1-\frac{|q|^2}{2})e^{-\frac{|q|^2}{4}}$. The second order corrections are  given by the following,
\begin{align}
   \Sigma_{1F}^{(BCS)} &= \sum_{m} 2(-1)^m \delta V^{\uparrow\uparrow,(BCS)}_{n=1,m} \approx -0.3688\\
   \Sigma_{1F}^{(EX)} &= \sum_{m} 2(-1)^m \delta V^{\uparrow\uparrow,(EX)}_{n=1,m} \approx -0.0002\\
   \Sigma_{1F}^{(VC)} &= \sum_{m} 2(-1)^m \delta V^{\uparrow\uparrow,(VC)}_{n=1,m} \approx 0.3878\\
   \Sigma_{1F}^{(BB)} &= \sum_{m} 2(-1)^m \delta V^{\uparrow\uparrow,(BB)}_{n=1,m} \approx -0.6244
\end{align}

\subsection{Correction to The Opposite-spin Haldane pseudopotential in $n=1$ LL}

In this section, we compute the second-order correction to the opposite-spin Haldane pseudopotentials. We restrict our analysis to the BCS and exchange diagram contributions, since the bubble and vertex corrections are identical for like-spin and opposite-spin pseudopotentials. The reasons are the following.

For vertex correction, the matrix elements of the like-spin and opposite-spin Coulomb interactions are given by:
\begin{align}
    \delta V^{\uparrow\uparrow,VC}_{12,34} &= \sum_{5,6}V^{\uparrow\uparrow}_{16,35} \frac{\pi(n_5,n_6)}{n_5-n_6}V^{\uparrow\uparrow}_{25,64}\\
        \delta V^{\uparrow\downarrow,VC}_{1\bar{2},3\bar{4}} &= \sum_{5,6}V^{\uparrow\downarrow}_{1\bar{6},3\bar{5}} \frac{\pi(n_5,n_6)}{n_5-n_6}V^{\uparrow\uparrow}_{25,64}\\
        & = \sum_{5,6}V^{\uparrow\uparrow}_{15,36} \frac{\pi(n_5,n_6)}{n_5-n_6}V^{\uparrow\uparrow}_{25,64}
\end{align}

For the opposite-spin vertex correction matrix elements, we  used the property in Eq.~\eqref{eq:Coulomb_symmetric}.  By relabeling indices $5$ and $6$ in the previous equation, we can see that$
\delta V^{\uparrow\uparrow, \mathrm{VC}}_{14,32} = \delta V^{\uparrow\downarrow, \mathrm{VC}}_{1\bar{2},3\bar{4}}
$. This is the same equation satisfied by the bare Coulomb interaction in Eq.~\eqref{eq:Coulomb_symmetric}. As a result, the vertex correction to like-spin and opposite-spin Haldane pseudo-potential are the same,
$
\delta V^{\uparrow\uparrow, \mathrm{VC}}_{m,n=1} = \delta V^{\uparrow\downarrow, \mathrm{VC}}_{m,n=1}
$. 
Similarly, the matrix elements of bubble diagram for like-spin and opposite-spin Coulomb interaction are given by:
\begin{align}
\delta V^{\uparrow\downarrow,BB}_{1\bar{2},3\bar{4}} &=- 2\sum_{5,6} V_{1\bar{5},3\bar{6}}^{\uparrow\downarrow}\frac{\pi(n_5,n_6)}{n_5-n_6}V_{62,54}^{\uparrow\uparrow} -2\sum_{5,6} V_{15,36}^{\uparrow\uparrow}\frac{\pi(n_5,n_6)}{n_5-n_6}V_{\bar{6}2,\bar{5}4}^{\uparrow\downarrow}\\
& = - 2\sum_{5,6} V_{16,35}^{\uparrow\uparrow}\frac{\pi(n_5,n_6)}{n_5-n_6}V_{62,54}^{\uparrow\uparrow} -2\sum_{5,6} V_{15,36}^{\uparrow\uparrow}\frac{\pi(n_5,n_6)}{n_5-n_6}V_{52,64}^{\uparrow\uparrow}\\
\delta V^{\uparrow\uparrow,BB}_{12,34} &=- 2\sum_{5,6} V_{15,36}^{\uparrow\uparrow}\frac{\pi(n_5,n_6)}{n_5-n_6}V_{62,54}^{\uparrow\uparrow} -2\sum_{5,6} V_{1\bar{5},3\bar{6}}^{\uparrow\downarrow}\frac{\pi(n_5,n_6)}{n_5-n_6}V_{\bar{6},2\bar{5}4}^{\uparrow\downarrow}\\
& = - 2\sum_{5,6} V_{15,36}^{\uparrow\uparrow}\frac{\pi(n_5,n_6)}{n_5-n_6}V_{62,54}^{\uparrow\uparrow} -2\sum_{5,6} V_{16,35}^{\uparrow\uparrow}\frac{\pi(n_5,n_6)}{n_5-n_6}V_{52,64}^{\uparrow\uparrow}
\end{align}
We  see that the matrix elements also satisfy $\delta V^{\uparrow\downarrow,BB}_{1\bar{4},3\bar{2}} = \delta V^{\uparrow\uparrow,BB}_{12,34}$, demonstrating that the bubble diagram does not distinguish between like-spin and opposite-spin Haldane pseudopotentials. Therefore, in the following we restrict our analysis of the opposite-spin case to the BCS and exchange diagram contributions.

\paragraph{BCS diagram}
Here we evaluate the BCS contribution to the opposite-spin Haldane pseudo-potential by making an analogy with the Exchange contribution to the like-spin, the reason is the following.
First, 
the BCS diagram for the opposite-spin is obtained by setting the $\sigma=\uparrow,\sigma'=\downarrow$ in Eq.~\eqref{eq:BCS_diagram}, its matrix elements are given by:

\begin{align}
    \delta V^{\uparrow\downarrow,BCS}_{1\bar{2},3\bar{4}} &= \sum_{5,6} V_{1\bar2,5\bar6}^{\uparrow\downarrow}\frac{\theta(n_5,n_6)}{2-n_5-n_6}V_{5\bar6,3\bar4}^{\uparrow\downarrow}-\sum_{m_5,m_6,n_5=n_6=0} V_{1\bar2,5\bar6}^{\uparrow\downarrow}\frac{1}{2}V_{5\bar6,3\bar4}^{\uparrow\downarrow}\\
    & = \sum_{5,6} V_{16,52}^{\uparrow\uparrow}\frac{\theta(n_5,n_6)}{2-n_5-n_6}V_{54,36}^{\uparrow\uparrow} - \sum_{m_5,m_6} V_{16,52}^{\uparrow\uparrow}\frac{1}{2}V_{54,36}^{\uparrow\uparrow} \\
    & = \delta V^{\uparrow\downarrow,BCS}_{14,32}
\end{align}

From the above expression, we again used Eq.~\eqref{eq:Coulomb_symmetric} to get rid of the bar and the $4$ and $2$ is now interpreted as hole state with opposite spin than the particle state $1$ and $3$.  
Noting that, for the like-spin exchange diagram, the matrix elements are given by(see Eq.~\eqref{eq:first quantization}, to compare with the BCS opposite-spin, we relabel the $m_2$ and $m_4$, this is different from using the properties of the wavefunction and get rid of the bar that we see previously):

\begin{align}
    \delta V^{\uparrow\uparrow,EX}_{14,32} =\langle 1\uparrow,4\uparrow|\delta V^{\uparrow\uparrow,EX}|3\uparrow,2\uparrow\rangle=\sum_{5,6}V^{\uparrow\uparrow}_{16,52}\frac{\pi(n_5,n_6)}{n_5-n_6}V^{\uparrow\uparrow}_{54,36}
\end{align}
The matrix elements of the like-spin exchange diagram closely resemble those of the 
opposite-spin BCS diagram, differing only in the energy denominator and the summation 
constraints imposed by \(\theta\) and \(\pi\). 
The exchange correction to the like-spin pseudopotential is obtained by transforming 
the two-particle states \((m_1,m_4)\) and \((m_3,m_2)\) into two-particle COM and relative states. 
In contrast, the BCS correction to the opposite-spin pseudopotential is obtained by transforming 
the two-body particle--hole state \((m_1,m_4)\) into particle--hole COM and relative states. 
Therefore, the calculation of \(\delta V_{n=1,m}^{\uparrow\downarrow,\mathrm{BCS}}\) proceeds 
in direct analogy to that of \(\delta V_{n=1,m}^{\uparrow\uparrow,\mathrm{EX}}\). 
Repeating the same steps as in the \(\delta V_{n=1,m}^{\uparrow\uparrow,\mathrm{EX}}\) calculation, 
the BCS contribution to the opposite-spin Haldane pseudopotential, 
\(\delta V^{\uparrow\downarrow,\mathrm{BCS}}_{m,n=1}\), is given by:

\begin{align}
   \delta V^{\uparrow\downarrow, \mathrm{BCS}}_{m,n=1}
    &=- \int \int \frac{dq_1^2dq_2^2}{(2\pi)^4}V_{q_1}V_{q_2} e^{-|q_1+q_2|^2/2}L_m(|q_1+q_2|^2)  \sum_{n_5,n_6}F_{1n_5}(q_1)F_{n_51}(q_2)  \frac{\theta(n_5,n_6)}{n_5+n_6-2}  F_{1n_6}(-q_2)F_{n_61}(-q_1) \notag \\
    &-\frac{1}{2}\int \int \frac{dq_1^2dq_2^2}{(2\pi)^4}V_{q_1}V_{q_2} e^{-|q_1+q_2|^2/2}L_m(|q_1+q_2|^2)  F_{10}(q_1)F_{01}(q_2) F_{10}(-q_2)F_{01}(-q_1) \label{eq:pp and hh opposite}
\end{align}

The first term corresponds to the virtual particle--particle excitation, 
while the second term represents the virtual hole--hole excitation. 
In contrast to the like-spin case, the virtual hole--hole process also contributes 
to the \(m=0\) and \(m=1\) opposite-spin Haldane pseudopotentials. Evaluating the above expression gives the BCS correction to the opposite-spin Haldane pseudopotential in the Table~I in main text.

\paragraph{Exchange diagram}
The exchange-diagram correction to the opposite-spin Haldane pseudopotentials 
can likewise be obtained by analogy with the BCS correction to the like-spin case. 
To this end, we first compare the matrix elements of the like-spin BCS diagram 
with those of the opposite-spin exchange diagram:

\begin{align}
    \delta V^{\uparrow\downarrow,EX}_{1\bar{2}3\bar{4}} 
    &=\sum_{5,6}V^{\uparrow\downarrow}_{1\bar6,5\bar4}\frac{\pi(n_5,n_6)}{n_5-n_6}V^{\uparrow\downarrow}_{5\bar2,3\bar6}=  \sum_{5,6} V_{1456}^{\uparrow\uparrow}\frac{\pi(n_5,n_6)}{n_5-n_6}V_{5632}^{\uparrow\uparrow}\\
    \delta V^{\uparrow\uparrow,BCS}_{1432} &=\sum_{5,6} V_{1456}^{\uparrow\uparrow}\frac{\theta(n_5,n_6)}{n_5+n_6-2}V_{5632}^{\uparrow\uparrow}
\end{align}
We find that the matrix elements are again similar, differing only in the energy denominator 
and in the summation range constrained by \(\theta\) and \(\pi\). 
For the same reasons as in the opposite-spin BCS correction, and by repeating the steps used 
in the like-spin BCS calculation, we obtain the following expression for the exchange correction 
to the opposite-spin Haldane pseudopotentials, 
\(\delta V^{\uparrow\downarrow,\mathrm{EX}}_{m,n=1}\):
\begin{align}
   \delta V^{\uparrow\downarrow,EX}_{m,n=1} &= {}_{(1)}\langle m| \delta V^{\uparrow\downarrow,EX}|m \rangle_{(1)} \\
   &=\sum_{5,6} {}_{(1)}\langle m|V|5,6\rangle \frac{\pi(n_5,n_6)}{n_5-n_6}\langle 5,6|V|m\rangle_{(1)}\\
   & =\sum_{n_5n_6,J,j}\sum_{n,n'} {}_{(1)}\langle m|V|n_5+n_6-n,J\rangle_c|n,j\rangle_r R^{n_5+n_6}_{n,n_6}\frac{\pi(n_5,n_6)}{n_5-n_6} {}_c\langle n_5+n_6-n',J|{}_r\langle n',j|V|m\rangle_{(1)}R^{n_5+n_6}_{n',n_6} \label{eq:ph_ex_opposite}
\end{align}
As in the like-spin BCS case, the expression yields four contributions---two square terms 
and two identical cross terms---so that 
\(\delta V^{\uparrow\downarrow,\mathrm{EX}}_{m,n=1} 
= \delta V^{\uparrow\downarrow,\mathrm{EX}}_1 
+ 2\,\delta V^{\uparrow\downarrow,\mathrm{EX}}_2 
+ \delta V^{\uparrow\downarrow,\mathrm{EX}}_3\). 
The first term, \(\delta V^{\uparrow\downarrow,\mathrm{EX}}_1\), is obtained by following the same steps 
as in the like-spin BCS calculation:

\begin{align}
  \delta V^{\uparrow\downarrow,EX}_1 & = \frac{1}{2}\sum_{n_5,n_6}\sum_{n}^{n_5+n_6} \big|{}_r\langle 0,m|V|n,n+m\rangle_r\big|^2 
  \delta_{2,n_5+n_6-n}\frac{\pi(n_5,n_6)}{n_5-n_6}  (R^{n_5+n_6}_{n,n_6})^2
\end{align}
The only difference of the above expression with BCS like-spin is the summation range and energy denominator. We can further simplify the above expression; Since $\pi(n_5,n_6)$ restricts $n5\geq1$ and $n_6=0$, let's relabel $n_5=\alpha$, thus we get:

\begin{align}
    \delta V^{\uparrow\downarrow,EX}_1 &=  \frac{1}{2}\sum_{\alpha\geq1}\sum_{n}^{\alpha} \big|{}_r\langle 0,m|V|n,n+m\rangle_r\big|^2 \frac{\delta_{2,\alpha-n}}{\alpha}  \left((R^{\alpha}_{n,0})^2\right)\\
    &=\frac{1}{2} \sum_{n\geq0} \big|{}_r\langle 0,m|V|n,n+m\rangle_r\big|^2 \frac{1}{n+2}  \frac{(n+1)(n+2)}{2^{n+3}}\\
    &=\frac{1}{2} \sum_{n\geq0} \big|{}_r\langle 0,m|V|n,n+m\rangle_r\big|^2   \frac{(n+1)}{2^{n+3}}
\end{align}

The cross term $\delta V^{\uparrow\downarrow,EX}_2$ is calculated following the same steps as $\delta V^{\uparrow\downarrow,EX}_1$:

\begin{align}
    \delta V^{\uparrow\downarrow,EX}_2
   & =  -\frac{1}{2}\sum_{n_5n_6}\sum_{n=0}^{n_5+n_6} {}_r\langle 0,m|V|n,n+m\rangle_r \delta_{2,n_5+n_6-n}R^{n_5+n_6}_{n,n_6}  \frac{\pi(n_5,n_6)}{n_5-n_6} {}_r\langle n+2,n+m|V|2,m\rangle_r R^{n_5+n_6}_{n+2,n_6}\\
   & = -\frac{1}{2}\sum_{\alpha\geq1}\sum_{n=0}^{\alpha} {}_r\langle 0,m|V|n,n+m\rangle_r \frac{\delta_{2,\alpha-n}}{\alpha}{}_r\langle n+2,n+m|V|2,m\rangle_r R^{\alpha}_{n+2,0}R^{\alpha}_{n,0}\\
& = -\frac{1}{2}\sum_{n\geq0} {}_r\langle 0,m|V|n,n+m\rangle_r {}_r\langle n+2,n+m|V|2,m\rangle_r  \frac{1}{n+2}R^{n+2}_{n+2,0}R^{n+2}_{n,0}\\
& =- \frac{1}{2} \sum_{n\geq0} {}_r\langle 0,m|V|n,n+m\rangle_r {}_r\langle n+2,n+m|V|2,m\rangle_r  \frac{1}{2^{n+2}}\sqrt{\frac{n+1}{2(n+2)}}
\end{align}

Finally ,the third term is given by:
\begin{align}
     \delta V^{\uparrow\downarrow,EX}_3
   & =  \frac{1}{2}\sum_{n_5n_6}\sum_{n=0}^{n_5+n_6} \big|{}_r\langle 2,m|V|n,n+m-2\rangle_r\big|^2\delta_{0,n_5+n_6-n}(R^{n_5+n_6}_{n,n_6})^2\frac{\pi(n_5,n_6)}{n_5-n_6}\\
   & = \frac{1}{2}\sum_{\alpha\geq1}\sum_{n=0}^{\alpha} \big|{}_r\langle 2,m|V|n,n+m-2\rangle_r\big|^2 \delta_{0,\alpha-n}\frac{(R^{\alpha}_{n,0})^2}{\alpha}= \frac{1}{2}\sum_{n\geq1}\big|{}_r\langle 2,m|V|n,n+m-2\rangle_r\big|^2\frac{1}{2^nn}
\end{align}

So the total exchange correction is given by (note that there's another identical contribution from exchange of $5$ and $6$):
\begin{align}
   & \delta V^{\uparrow\downarrow,EX}_{n=1,m} \notag \\
   &= 2\left(\delta V^{\uparrow\downarrow,EX}_1+2\delta V^{\uparrow\downarrow,EX}_2+\delta V^{\uparrow\downarrow,EX}_3\right) = \sum_{n\geq0} \big|{}_r\langle 0,m|V|n,n+m\rangle_r\big|^2   \frac{(n+1)}{2^{n+3}}+\sum_{n\geq1}\big|{}_r\langle 2,m|V|n,n+m-2\rangle_r\big|^2\frac{1}{2^nn}\notag \\
    &- 2\sum_{n\geq0} {}_r\langle 0,m|V|n,n+m\rangle_r {}_r\langle n+2,n+m|V|2,m\rangle_r \frac{1}{2^{n+2}}\sqrt{\frac{n+1}{2(n+2)}}\\
    & = \sum_{n\geq0}\frac{1}{2^{n+1}}\bigg( \big|{}_r\langle 0,m|V|n,n+m\rangle_r\big|^2   \frac{(n+1)}{4}+ \big|{}_r\langle 2,m|V|n+1,n+m-1\rangle_r\big|^2\frac{1}{n+1} \notag  \\
    &- {}_r\langle 0,m|V|n,n+m\rangle_r {}_r\langle n+2,n+m|V|2,m\rangle_r \sqrt{\frac{n+1}{2(n+2)}} \bigg)
\end{align}
Evaluating the above expression numerically generates the exchange diagram to the opposite-spin Haldane pseudopotentials present in the main text Table~I.
Together with both the like-spin and opposite-spin Haldane pseudopotentials and their corrections, the results are  summarized in the table below.
\begin{table*}[b]
    \centering
    \renewcommand{\arraystretch}{1.2}
    \begin{tabular}{|c|cc|cc|cc|cc|cc|c|c|}
        \hline
        Haldane Pseudopotentials & \multicolumn{2}{c|}{$m=0$} & \multicolumn{2}{c|}{$m=1$} & \multicolumn{2}{c|}{$m=2$} & \multicolumn{2}{c|}{$m=3$} & \multicolumn{2}{c|}{$m=4$} & \multicolumn{2}{c|}{Fock Self Energy $\Sigma_F^{(\alpha)}$} \\
        \cline{2-13}
        & $\uparrow\uparrow$ & $\uparrow\downarrow$ & $\uparrow\uparrow$ & $\uparrow\downarrow$
        & $\uparrow\uparrow$ & $\uparrow\downarrow$ & $\uparrow\uparrow$ & $\uparrow\downarrow$
        & $\uparrow\uparrow$ & $\uparrow\downarrow$ & Bare Value &  Correction     \\
        \hline
        \multicolumn{13}{|c|}{\textbf{First Landau level (\(n=1\))}} \\
        \hline
        $V_{n=1,m}$              & \multicolumn{2}{c|}{0.6093} & \multicolumn{2}{c|}{0.4154}
                                 & \multicolumn{2}{c|}{0.4500} & \multicolumn{2}{c|}{0.3150}
                                 & \multicolumn{2}{c|}{0.2635} &  &  \\
        $\delta V^{\mathrm{BB}}_{n=1,m}$
                                 & -0.3930 & -0.3930 & -0.2038 & -0.2038
                                 & -0.1981 & -0.1981 & -0.1119 & -0.1119
                                 & -0.0535 & -0.0535 &  & -0.6244 \\
        $\delta V^{\mathrm{VC}}_{n=1,m}$
                                 &  0.0750 &  0.0750 & -0.0475 & -0.0475
                                 &  0.0870 &  0.0870 &  0.0141 &  0.0141
                                 & -0.0128 & -0.0128 &  & 0.3878 \\
        $\delta V^{\mathrm{EX}}_{n=1,m}$
                                 &  0.0247 &  0.0750 &  0.0706 &  0.0135
                                 &  0.0803 &  0.0841 &  0.0186 &  0.0152
                                 & -0.0031 &  0.0066 & $\frac{3}{4}\sqrt{\frac{\pi}{2}}$ & -0.0002 \\
        $\delta V^{\mathrm{BCS}}_{n=1,m}$
                                 & -0.0890 & -0.1550 & -0.0347 &  0.0015
                                 & -0.1498 & -0.1091 & -0.0264 & -0.0329
                                 & -0.0116 & -0.0151 &  & -0.3688 \\
        $\sum_{\alpha}\delta V^{\alpha}_{n=1,m}$
                                 & -0.3823 & -0.3980 & -0.2154 & -0.2363
                                 & -0.1806 & -0.1361 & -0.1056 & -0.1155
                                 & -0.0810 & -0.0748 &  & -0.6056 \\
        \hline
    \end{tabular}
    \caption{Spin-resolved Haldane pseudopotentials in the first (\(n=1\)) Landau levels. Each \(m\)-sector has two columns corresponding to like-spin (\(\uparrow\uparrow\)) and opposite-spin (\(\uparrow\downarrow\)) Coulomb interactions.  The first row gives the bare projected interaction, followed by second-order corrections from the indicated diagrammatic processes. In the last column, we show the second order correction to the self energy arising from those diagrams.}
    \label{tab:haldane_pseudopotentials_n1}
\end{table*}

\newpage
\subsection{Comparison of Equal-Spin and Opposite-Spin Pseudopotentials in the $n=1$ Landau level}
In this subsection, we provide a detailed comparison between the equal-spin and opposite-spin Haldane pseudopotentials in the $n=1$ LL and identify the microscopic origin of the spin anisotropy. The anisotropy arises solely from the BCS and double-exchange diagrams. While the equal-spin and opposite-spin channels share the same phase space for virtual transitions, the matrix elements between states in the $n=1$ LL manifold and the virtual states are different. The difference in the matrix elements is the microscopic origin of the spin anisotropy.
\begin{figure}[h]
    \centering
    \includegraphics[width=1\linewidth]{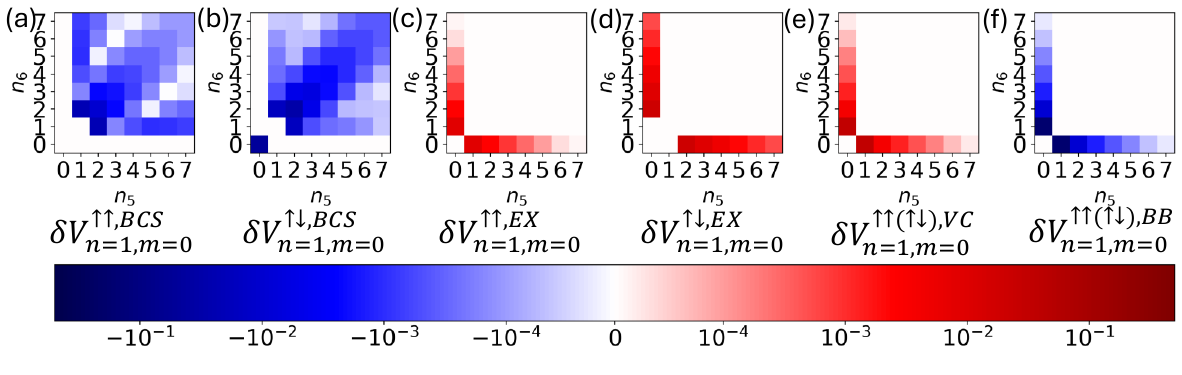}
    \caption{The graphs illustrate the screening difference between the like-spin and opposite-spin Coulomb interactions for $m=0$ component at $n=1$ LL. The horizontal and vertical axes correspond to the Landau level indices of the virtual excitation states, denoted by $n_{5}$ and $n_{6}$, respectively, in the expression for the second-order correction.}
    \label{fig:virtual_state_contribution}
\end{figure}

\subsubsection{$n=1$ Landau level, $m=0$ component}

We begin our discussion with the $m=0$ Haldane pseudopotential. Figure~\ref{fig:virtual_state_contribution} shows the contributions from the four diagrams to the $m=0$ pseudopotential for both equal-spin and opposite-spin. The $x$–$y$ in each subplot correspond to the Landau-level indices of the virtual excited states. Panels (a)–(b) correspond to the BCS diagrams [Eqs.~\eqref{eq:pp and hh}, \eqref{eq:pp and hh opposite}], panels (c)–(d) to the double-exchange diagrams [Eqs.~\eqref{eq:ph_ex}, \eqref{eq:ph_ex_opposite}], panel (e) to the vertex correction [Eq.~\eqref{eq:ph_vc_like}], and panel (f) to the bubble diagram [Eq.~\eqref{eq:ph_bb_like}].  Note that the virtual excitation phase space for the BCS process (two particles or two holes) is much larger than that for the particle–hole diagrams.

We first analyze the BCS diagram. As shown in Table~2, there is a small spin anisotropy in this angular momentum: $\delta V_{n=1,m=0}^{\uparrow\downarrow,\mathrm{BCS}} = -0.1550$ is slightly smaller than $\delta V_{n=1,m=0}^{\uparrow\uparrow,\mathrm{BCS}} = -0.0890$. This difference can be understood from Fig.~\ref{fig:virtual_state_contribution}a) and b). First, we note that all contribution is negative . For the like-spin case, certain excited state contribution is exactly zero like $(n_5,n_6)=(0,0)$ and $(n_5,n_6)=(5,2)$ and $(n_5,n_6)=(2,5)$. This is because of the selection rules impose by angular momentum conservation. The two-particle state in $n=1$ LL with COM guiding center $M$ and relative guiding center $m$ is an equal superposition of the following two COM LL and relative LL basis
\begin{align}
   |n=1,n=1\rangle |M\rangle_c|m\rangle_r &= \frac{1}{\sqrt{2}}\big( |N=2,M\rangle_c|n=0,m\rangle_r-|N=0,M\rangle_c|n=2,m\rangle_r \big) 
\end{align}
Here the subscript $c$ and $r$ represent COM and relative coordinate. The matrix elements with $n_5$ and $n_6$
\begin{align}
    \langle n_5m_5, n_6m_6|V|n=1,n=1\rangle |M\rangle_c|m=0\rangle_r &=\sum_{N',n',J,j}R^{n_5+n_6}_{n',n_6}R^{m_5+m_6}_{j,m_6}\delta_{N'+n',n_5+n_6}\delta_{J+j,m_5+m_6} \notag\\
    &\times\frac{1}{\sqrt{2}}{}_c\langle N',J|{}_r\langle n',j|V|\big( |N=2,M\rangle_c|n=0,m\rangle_r-|N=0,M\rangle_c|n=2,m\rangle_r \big) \\
&=\sum_{N',n',J,j}R^{n_5+n_6}_{n',n_6}R^{m_5+m_6}_{j,m_6}\delta_{N'+n',n_5+n_6}\delta_{J+j,m_5+m_6} \notag\\
    &\times\frac{1}{\sqrt{2}}({}_r\langle n',j|V|0,m\rangle_r\delta_{N',2}\delta_{J,M}- {}_r\langle n',j|V|2,m\rangle_r\delta_{N',0}\delta_{J,M})\label{eq:angularmomentum}
\end{align}
where we used Eq.~\eqref{eq:COM basis transformation} to transform the single particle basis into the COM--relative basis. For the virtual hole-hole state, e.g $n_5=n_6=0$, $N'=n'=0$, only the second term survive, but since the Coulomb interaction conserve the angular momentum, the relative index has to satisfies $0-j=2-m$ and note $j\geq0$, only the $m\geq2$ components will have non-zero values.
Such angular momentum conversation restriction does not apply in the opposite-spin case since wavefunctions have opposite chirality.  As a result, the $\delta V_{n=1,m=0}^{\uparrow\downarrow,\mathrm{BCS}}$ is more negative than $\delta V_{n=1,m=0}^{\uparrow\uparrow,\mathrm{BCS}}$.

Next, we turn to the double-exchange diagrams shown in Figs.~\ref{fig:virtual_state_contribution}(c,d). In contrast to the BCS diagram, which reduces the strength of the Haldane pseudopotential, the double-exchange diagram enhances it: all BCS matrix elements are negative, whereas all double-exchange matrix elements are positive. As listed in Table~2, a small spin anisotropy is again present, with $\delta V_{n=1,m=0}^{\uparrow\downarrow,\mathrm{EX}} = 0.0750$ slightly larger than $\delta V_{n=1,m=0}^{\uparrow\uparrow,\mathrm{EX}} = 0.0247$. This difference also originates from matrix-element effects. In particular, the color map in Fig.~\ref{fig:virtual_state_contribution}(d) shows darker red regions compared with Fig.~\ref{fig:virtual_state_contribution}(c), even though contributions from $(n_5,n_6) = (1,0)$ or $(0,1)$ vanish due to angular-momentum selection rules.  

Taken together, the BCS diagram reduces the $m=0$ pseudopotential while the double-exchange diagram enhances it. Their competition is responsible for the net spin anisotropy in the $m=0$ Haldane pseudopotential.

%

Lastly, let's discuss the vertex-correction and bubble diagrams. They do not give rise to any spin anisotropy. 
The virtual-state contributions are identical for the like-spin and 
opposite-spin channels.  Physically, the bubble diagram generates charge-density fluctuations and therefore cannot produce spin anisotropy. In the vertex-correction diagram, one particle line is dressed by a spin-independent exchange process, which likewise preserves spin symmetry.

As shown in Table~II, the bubble diagram provides the dominant correction among all diagrams, even surpassing the BCS process despite the latter having a larger phase space. This is illustrated in the color maps of Fig.~\ref{fig:virtual_state_contribution}(a,b,f). Although the BCS diagram has a bigger range of phase space ($n_5+n_6\geq3$), contributions from states far from the projected LL ($n=1$) are strongly suppressed by the combined effect of matrix-element decay and large energy denominators. In contrast, although the bubble (particle-hole) diagram has smaller phase space, $(n_5=0, n_6\geq)1$ and $(n_6=0, n_5\geq1)$], it receives substantial weight from nearby LL, particularly from $(n_5,n_6)=(1,0)$ and $(n_5,n_6)=(0,1)$, making it the dominant contribution. The color maps shown in Fig.~\ref{fig:virtual_state_contribution}f) already included the spin degeneracy.

\subsubsection{$n=1$ Landau level, $m=1$ component}

\begin{figure}[h]
    \centering
    \includegraphics[width=1\linewidth]{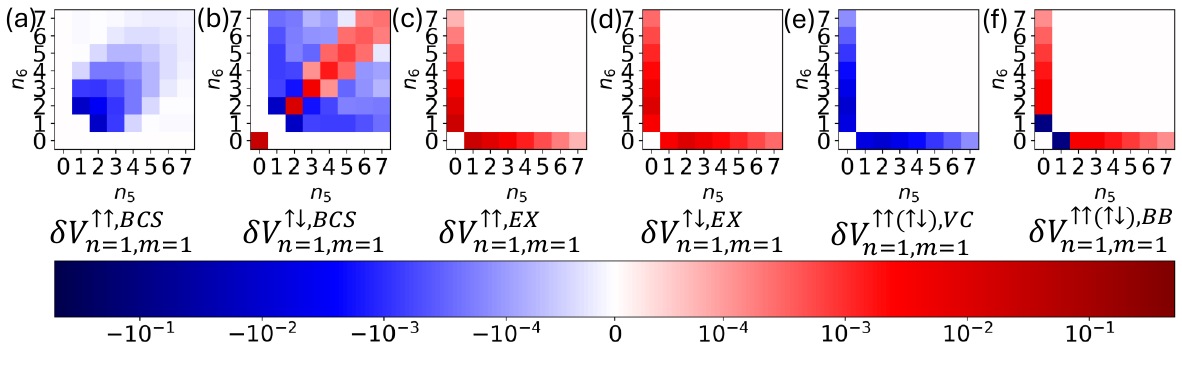}
    \caption{These graphs are analogous to Fig.~\ref{fig:virtual_state_contribution}, 
but correspond to the $m=1$ Haldane pseudo-potentials. 
}
    \label{fig:virtual_state_contribution_m1}
\end{figure}

Next, we discuss the $m=1$ case. Figure~\ref{fig:virtual_state_contribution_m1} shows the BCS diagrams in panels (a)--(b), the exchange diagrams in (c)--(d), the vertex correction in (e), and the bubble diagram in (f).

We begin with the BCS diagrams again.  As shown in Table.~\ref{tab:haldane_pseudopotentials_n1}, the BCS diagram reduce the like-spin but enhance the opposite-spin Haldane pseudopotentials, with corrections given by $\delta V_{n=1,m=1}^{\uparrow\uparrow}=-0.0347$ while $\delta V_{n=1,m=1}^{\uparrow\uparrow}=0.0015$.
The like-spin channel does not 
include the hole--hole contribution, due to the angular momentum conservation in Eq.~\eqref{eq:angularmomentum} and the individual 
contribution decrease with increasing component index, resulting in only a weak 
screening effect at higher components. The opposite-spin channel  receive positive contribution from certain virtual states, like the hole--hole state(e.g$(n_5=0,n_6=0)$) and some particle--particle states (e.g $(n_5=2,n_6=2)$), 
this ultimately leads to an overall positive BCS correction to the opposite-spin 
Haldane pseudo-potential. Thus, the BCS diagrams reduce the strength of the 
$m=1$ like-spin pseudo-potential, while enhancing that of the opposite-spin channel.  

Turning to the exchange diagrams, both the like-spin and opposite-spin channels are enhanced. Since virtual states with smaller energy denominators dominate the screening effect, and the smallest-denominator contribution in the like-spin channel is larger than that in the opposite-spin case, the exchange diagrams enhance the like-spin Haldane pseudopotential more strongly. In particular, we find $\delta V_{n=1,m=1}^{\uparrow\uparrow,EX}=0.0706$ and $\delta V_{n=1,m=1}^{\uparrow\downarrow,EX}=0.0135$.

The vertex-correction diagrams remain 
identical between the two spin sectors and yield negative contributions.
As noted above, the bubble diagram dominates the overall screening effect, 
producing a large negative correction. This is evident in 
Fig.~\ref{fig:virtual_state_contribution_m1}(f), where the virtual states  $(n_5=1,n_6=0)$ and $(n_5=0,n_6=1)$
contributes a strong negative value that overwhelms the small positive 
contributions from other states.  

The net outcome of all these diagrams is that Landau-level mixing screens 
the opposite-spin channel more strongly than the like-spin channel. 
For higher components, we do not present explicit results here, as they 
follow directly from the reasoning outlined above.

\section{Derivation of the RPA Matrix} We derive here the RPA matrix $A$ used in the main text. The QHF state is defined as the one in which the majority-spin LL ($\downarrow$) is completely filled while the minority-spin LL ($\uparrow$) is empty:
\begin{align}
|QHF_n\rangle = \prod_{m} c^\dagger_{nm\downarrow}|0\rangle .
\end{align}
This is an eigenstate of the LL-projected Hamiltonian $H_n$ discussed in the main text. At leading order in the LL-mixing parameter $\kappa$, it is the ground state (lowest-energy eigenstate). However, once LL mixing at order $\kappa^2$ is included, $|QHF_n\rangle$ remains an eigenstate but is not guaranteed to be the ground state, since the Hamiltonian no longer has a manifestly non-negative spectrum.

Eigenstates of the $n$th LL-projected Hamiltonian can be labeled by their $S_z$ quantum number. A single spin-reversal excitation (a spin exciton) lies in the sector $S_z = N/2 - 1$, where $N = N_\phi$ is the number of electrons (or flux quanta). This Hilbert space has dimension $N^2$, because one can remove any of the $N$ occupied majority-spin states and place it into any of the $N$ empty minority-spin states. The basis states of this sector are
\begin{align}
|m_1\uparrow,m_2\downarrow\rangle = Q^\dagger_{n,m_1m_2}|QHF_n\rangle ,
\end{align}
where the spin-exciton creation operator is
\begin{align}
Q^\dagger_{n,m_1m_2} = c^\dagger_{nm_1\uparrow} c_{nm_2\downarrow}.
\end{align}

Two key properties are worth noting:
\begin{enumerate}
\item Because opposite-spin LLs experience opposite effective magnetic fields, the particle–hole pair $|m_1\uparrow,m_2\downarrow\rangle$ has the same chirality (no bar on top of $m$).
\item The QHF is annihilated by the exciton annihilation operator because it has the maximum spin-polarization
\begin{align}
Q_{n,m_1m_2}|QHF_n\rangle = 0 .
\end{align}
\end{enumerate}

A general state in this Hilbert space is
\begin{align}
|\psi_{na}\rangle = \sum_{m_1,m_2} \chi^{a}_{n,m_1m_2} Q^\dagger_{n,m_1m_2}|QHF_n\rangle ,
\end{align}
and satisfies the eigenvalue problem
\begin{align}
\mathcal{H}^{\text{eff}}_n |\psi_{na}\rangle = E_{na}|\psi_{na}\rangle .
\end{align}
The QHF itself satisfies
\begin{align}
\mathcal{H}^{\text{eff}}_n|QHF_n\rangle = E_{n0}|QHF_n\rangle .
\end{align}
When $E_{na}<E_{n0}$, the exciton has lower energy than the QHF, signaling an instability of QHF. To derive the RPA equation, using $Q_{n,m_1m_2}|QHF_n\rangle=0$, we obtain
\begin{align}
\sum_{m_1,m_2}\chi^{a}_{n,m_1m_2} [\mathcal{H}^{\text{eff}}_n, Q^\dagger_{n,m_1m_2}]|QHF_n\rangle
&= (E_{na}-E_{n0})\sum_{m_1,m_2}\chi^{a}_{n,m_1m_2} Q^\dagger_{n,m_1m_2}|QHF_n\rangle .
\end{align}
Acting further with $Q_{n,m_3m_4}$ and using $Q|QHF_n\rangle=0$, the RPA equation becomes the famous double-commutator form,
\begin{align}
\sum_{m_1,m_2}\chi^{a}_{n,m_1m_2},[Q_{n,m_3m_4},[\mathcal{H}^{\text{eff}}_n,Q^\dagger_{n,m_1m_2}]]|QHF_n\rangle
= (E_{na}-E_{n0}) \sum_{m_1,m_2}\chi^{a}_{n,m_1m_2} [Q_{n,m_3m_4},Q^\dagger_{n,m_1m_2}]|QHF_n\rangle .
\end{align}
Multiplying on the left by $\langle QHF_n|$ yields the RPA eigenvalue problem,
\begin{align}
\sum_{m_1,m_2} A_{n;m_1m_2,m_3m_4} \chi^{a}_{n,m_1m_2} = (E_{na}-E_{n0})\chi^{a}_{n,m_3m_4},
\end{align}
where the RPA matrix is
\begin{align}
A_{n;m_1m_2,m_3m_4}
&= \big(\epsilon^{HF}_{nm_1\uparrow}-\epsilon^{HF}_{nm_2\downarrow}\big)\delta_{m_1m_3}\delta_{m_2m_4}
- V^{\uparrow\downarrow}_{n;m_1m_2,m_3m_4}
+ \kappa^2 \langle n m_1\uparrow, n m_2\downarrow| W_n |n m_3\uparrow, n m_4\downarrow \rangle .
\label{eq:RPA_matrix}
\end{align}
The first term is the renormalized Hartree–Fock self-energy,
\begin{align}
    \epsilon^{HF }_{nm_1\uparrow}-\epsilon^{HF}_{nm_2\downarrow} 
    = \Sigma_{n,F} = \sum_{m'}V^{\uparrow\uparrow}_{nm,nm';nm',nm}, = \sum_{m}2(-1)^m (\kappa V^{\uparrow\uparrow}_{n,m}+\kappa^2 \delta V^{\uparrow\uparrow}_{n,m})
\end{align}
which is independent of guiding center in a uniform liquid. This term contains both the bare and second order correction. The second term represents the renormalized opposite-spin Coulomb interaction and the third term is the three-body contribution discussed in the next section.

Since the exciton interaction depends only on the relative guiding centers and not on the center of mass, the RPA matrix $A$ can be block diagonalized in the center-of-mass guiding-center basis:
\begin{align}
    A_n = \sum_{m,M}A_{n,m} |M,m\rangle \langle M,m|.
\end{align}
The block eigenvalues take the form
\begin{align}
    A_{n,m} &= \Sigma_{n,F} -\mathcal{V}^{\uparrow\downarrow}_{n,m} +\kappa^2 W_{n,m}.\\
   \mathcal{V}^{\uparrow\downarrow}_{n,m} & = \kappa V_{n,m}^{\uparrow\downarrow}+\kappa^2 \delta V^{\uparrow\downarrow}_{n,m}.
\end{align}
This is the result presented in the main text.
Here the second term is the renormalized opposite-spin Haldane pseudopotential, and the third term encodes the three-body interaction that we will calculate in the following section.

\section{Three-body interaction correction to the spin-exciton energys at $n=1$}\label{sec:three_body_interaction}


The second-order perturbation of the Coulomb interaction also generates 
effective three-body terms, and as discussed in the main text, obtaining the exciton energy to exact second 
order requires inclusion of the three-body 
interaction. Now we will study its effect.

We begin with the three-body interaction obtained from second-order degenerate 
perturbation theory, projected onto the $n=1$ Landau level, e.g Eq.~\eqref{eq:three_body_n1}. The analysis for 
$n=0$ follows analogously and is omitted here since the only difference is that in the $n=0$, the energy denormiator is $n_7$ and the range is $n_7\geq1$ as shown in Eq.~\eqref{eq:three_body_n0}, so only the resulting contribution 
is reported in the main text. The expression of the three-body interaction is listed again below
:
\begin{align}
W_1 & =-\sum_{\sigma,\sigma',\sigma''}\sum_{1,2',3'',4'',5',6}\sum_{n_7\neq1,m_7} V^{\sigma\sigma'}_{1,2';7,5'}V^{\sigma\sigma''}_{7,3'';6,4''} \frac{1}{n_7-1}c_{1\sigma}^\dagger c_{2'\sigma'}^\dagger c_{3''\sigma''}^\dagger c_{4''\sigma''} c_{5'\sigma'}c_{6\sigma}\label{eq:W1}
\end{align}

The contribution to the spin-exciton energy from the three-body interaction is listed in the third term in Eq.~\eqref{eq:RPA_matrix}. Here we write it down again:

\begin{align}
    \langle  m_1\uparrow,  m_2\downarrow| W_1 | m_3\uparrow, m_4\downarrow \rangle = \langle QHF\downarrow|c_{m_2\downarrow}^\dagger c_{m_1\uparrow} W_1 c_{m_3\uparrow}^\dagger c_{m_4\downarrow}|QHF\downarrow \rangle
\end{align}
We omit the LL index since we restrict ourselves to the $n=1$ LL. The spin-exciton state is given by $c_{m_3\uparrow}^\dagger c_{m_4\downarrow}|QHF\downarrow \rangle$, which creates a spin-up particle in the empty spin-up LL and a hole in the fully filled spin-down LL. Thus, the state can be regarded as an effective two-body particle--hole pair. To make this structure explicit, we perform a particle--hole transformation on the spin-down LL.

\begin{align}
    c_{m\uparrow} \rightarrow c_m , \quad c_{m\downarrow} \rightarrow h_m^\dagger
\end{align}
where $h_m^\dagger$ denotes the hole creation operator. We omit the spin index here, since the particle operator always carries spin-up and the hole operator always carries spin-down. After the particle--hole transformation, both $c$ and $h$ annihilate the $|QHF\downarrow\rangle$ state:

\begin{align}
    c_{m}|QHF\downarrow\rangle =0, \quad h_{m}|QHF\downarrow\rangle=0
\end{align}
Thus, the state $|QHF\downarrow\rangle$ serves as the vacuum in the particle--hole transformed basis. In this representation, the spin-exciton state is written as $c_{m_3\uparrow}^\dagger c_{m_4\downarrow}|QHF\downarrow \rangle = c_{m_3}^\dagger h_{m_4}^\dagger|QHF\downarrow \rangle$ which is manifestly a two-body particle--hole pair. In what follows, we will work in this particle--hole basis to evaluate the three-body interaction correction. 

Since any normal-ordered three-body interaction acting on a two-body state vanishes, we normal order $W_1$ in the particle--hole basis by moving all annihilation operators ($c, h$) to the right and all creation operators ($c^\dagger, h^\dagger$) to the left in Eq.~\eqref{eq:W1}. This procedure generates additional two-body interaction terms as well as one-body energy shifts. The one-body shifts renormalize the single-particle energies of both particles and holes, thereby modifying the spin-exciton energy. The two-body interactions can be classified into three types: $V_{pp}$, $V_{ph}$, and $V_{hh}$, where $V_{pp}$ describes particle--particle interactions, $V_{hh}$ hole--hole interactions, and $V_{ph}$ particle--hole interactions. Among these, only $V_{ph}$ contributes to the exciton energy, since a single spin exciton consists of precisely one particle--hole pair.

In the following, we normal order $W_1$ in detail, focusing only on the one-body energy shift and the particle--hole interaction $V_{ph}$. To carry out the particle--hole transformation, we explicitly retain the spin dependence. In terms of the spin indices, this yields a total of eight terms and note that the state with different spin carries different chirality:

\begin{align}
   W_1 &= -\sum_{1,2,3,4,5,6}\sum_{7} V^{\uparrow\uparrow}_{1,2;7,5}V^{\uparrow\uparrow}_{7,3;6,4 } \frac{w(n_7)}{n_7-1}c_{1\uparrow}^\dagger c_{2\uparrow}^\dagger c_{3\uparrow}^\dagger c_{4\uparrow} c_{5\uparrow}c_{6\uparrow}-\sum_{1,2,3,4,5,6}\sum_{7} V^{\uparrow\uparrow}_{1,2;7,5}V^{\uparrow\downarrow}_{7,\bar3;6,\bar4 } \frac{w(n_7)}{n_7-1}c_{1\uparrow}^\dagger c_{2\uparrow}^\dagger c_{\bar3\downarrow}^\dagger c_{\bar4\downarrow} c_{5\uparrow}c_{6\uparrow}\\
    &-\sum_{1,2,3,4,5,6}\sum_{7} V^{\uparrow\downarrow}_{1,\bar2;7,\bar5}V^{\uparrow\uparrow}_{7,3;6,4 } \frac{w(n_7)}{n_7-1}c_{1\uparrow}^\dagger c_{\bar2\downarrow}^\dagger c_{3\uparrow}^\dagger c_{4\uparrow} c_{\bar5\downarrow}c_{6\uparrow}-\sum_{1,2,3,4,5,6}\sum_{7} V^{\uparrow\downarrow}_{\bar1,2;\bar7,5}V^{\uparrow\downarrow}_{\bar7,3;\bar6,4 } \frac{w(n_7)}{n_7-1}c_{\bar1\downarrow}^\dagger c_{2\uparrow}^\dagger c_{3\uparrow}^\dagger c_{4\uparrow} c_{5\uparrow}c_{\bar6\downarrow}\\
    &-\sum_{1,2,3,4,5,6}\sum_{7} V^{\uparrow\downarrow}_{\bar1,2;\bar7,5}V^{\uparrow\uparrow}_{7,3;6,4 } \frac{w(n_7)}{n_7-1}c_{\bar1\downarrow}^\dagger c_{2\uparrow}^\dagger c_{\bar3\downarrow}^\dagger c_{\bar4\downarrow} c_{5\uparrow}c_{\bar6\downarrow}-\sum_{1,2,3,4,5,6}\sum_{7} V^{\uparrow\uparrow}_{1,2;7,5}V^{\uparrow\uparrow}_{\bar7,3;\bar6,4 } \frac{w(n_7)}{n_7-1}c_{\bar1\downarrow}^\dagger c_{\bar2\downarrow}^\dagger c_{3\uparrow}^\dagger c_{4\uparrow} c_{\bar5\downarrow}c_{\bar6\downarrow}\\
    &-\sum_{1,2,3,4,5,6}\sum_{7} V^{\uparrow\downarrow}_{1,\bar2;7,\bar5}V^{\uparrow\downarrow}_{7,\bar3;6,\bar4 } \frac{w(n_7)}{n_7-1}c_{1\uparrow}^\dagger c_{\bar2\downarrow}^\dagger c_{\bar3\downarrow}^\dagger c_{\bar4\downarrow} c_{\bar5\downarrow}c_{6\uparrow}-\sum_{1,2,3,4,5,6}\sum_{7} V^{\uparrow\uparrow}_{1,2;7,5}V^{\uparrow\uparrow}_{7,3;6,4 } \frac{w(n_7)}{n_7-1}c_{1\downarrow}^\dagger c_{2\downarrow}^\dagger c_{3\downarrow}^\dagger c_{4\downarrow} c_{5\downarrow}c_{6\downarrow}
\end{align}
In the particle-hole basis, the three-body interaction becomes:
\begin{align}
    W_1 &= -\sum_{1,2,3,4,5,6}\sum_{7} V_{1,2;7,5}V_{7,3;6,4} \frac{w(n_7)}{n_7-1}c_{1}^\dagger c_{2}^\dagger c_{3}^\dagger c_{4} c_{5}c_{6}-\sum_{1,2,3,4,5,6}\sum_{7} V_{1,2;7,5}V_{7,4;6,3} \frac{w(n_7)}{n_7-1}c_{1}^\dagger c_{2}^\dagger h_{3} h_{4}^\dagger c_{5}c_{6}\\
    &-\sum_{1,2,3,4,5,6}\sum_{7} V_{1,5;7,2}V_{7,3;6,4} \frac{w(n_7)}{n_7-1}c_{1}^\dagger h_{2} c_{3}^\dagger c_{4} h_{5}^\dagger c_{6}-\sum_{1,2,3,4,5,6}\sum_{7} V_{7,2;1,5}V_{6,3;7,4} \frac{w(n_7)}{n_7-1}h_{1} c_{2}^\dagger c_{3}^\dagger c_{4} c_{5}h_{6}^\dagger\\
    &-\sum_{1,2,3,4,5,6}\sum_{7} V_{7,2;1,5}V_{6,4;7,3} \frac{w(n_7)}{n_7-1}h_{1} c_{2}^\dagger h_{3}h_{4}^\dagger c_{5}h_{6}^\dagger-\sum_{1,2,3,4,5,6}\sum_{7} V_{7,5;1,2}V_{6,3;7,4} \frac{w(n_7)}{n_7-1}h_{1} h_{2}c_{3}^\dagger c_{4} h_{5}^\dagger h_{6}^\dagger\\
    &-\sum_{1,2,3,4,5,6}\sum_{7} V_{1,5;7,2}V_{7,4;6,3} \frac{w(n_7)}{n_7-1}c_{1}^\dagger h_{2} h_{3} h_{4}^\dagger h^\dagger_{5}c_{6}-\sum_{1,2,3,4,5,6}\sum_{7} V_{7,5;1,2}V_{6,4;7,3} \frac{w(n_7)}{n_7-1}h_{1} h_{2} h_{3} h_{4}^\dagger h_{5}^\dagger h_{6}^\dagger
\end{align}
In the above expression we have used the property in Eq.~\eqref{eq:Coulomb_symmetric},  
$V^{\uparrow\downarrow}_{1,\bar{2};3,\bar{4}} = V^{\uparrow\uparrow}_{1,4;3,2} 
\equiv V_{1,4;3,2}$.  
Since both the particle and the hole experience the same magnetic field, no 
overbar is needed on the operator

The normal ordering of the first four terms in the expression above do not generate any one-body or two-body interactions.
For the fifth, sixth and seventh terms, it becoems the following:

\begin{align}
    h_{1} c_{2}^\dagger h_{3}h_{4}^\dagger c_{5}h_{6}^\dagger &= c_2^\dagger c_5\delta_{16}\delta_{34}-c_2^\dagger c_5 \delta_{36}\delta_{14}+c_2^\dagger h_6^\dagger h_3 c_5 \delta_{14}-c_2^\dagger h_6^\dagger h_1c_5\delta_{34}+c_2^\dagger h_4^\dagger h_1 c_5\delta_{36}-c_2^\dagger h_4^\dagger h_3 c_5\delta_{16}+c_2^\dagger h_4^\dagger h_6^\dagger h_1h_3c_5 \\
     h_{1} h_{2} c_{3}^\dagger c_{4} h_{5}^\dagger h_{6}^\dagger &= c_3^\dagger c_4\delta_{16}\delta_{25}-c_3^\dagger c_4 \delta_{26}\delta_{15}+c_3^\dagger h_6^\dagger h_2 c_4 \delta_{15}-c_3^\dagger h_6^\dagger h_1c_4\delta_{25}+c_3^\dagger h_5^\dagger h_1 c_4\delta_{26}-c_3^\dagger h_5^\dagger h_2 c_4\delta_{16}+c_3^\dagger h_5^\dagger h_6^\dagger h_1h_2c_4 \\
     c_{1}^\dagger h_{2}  h_{3}h_{4}^\dagger h_{5}^\dagger c_{6} &= c_1^\dagger c_6 \delta_{25} \delta_{34}
- c_1^\dagger c_6 \delta_{35} \delta_{24}
+ c_1^\dagger h_5^\dagger h_3 c_6 \delta_{24}
- c_1^\dagger h_5^\dagger h_2 c_6 \delta_{34}
 + c_1^\dagger h_4^\dagger h_2 c_6 \delta_{35}
- c_1^\dagger h_4^\dagger h_3 c_6 \delta_{25}
+ c_1^\dagger h_4^\dagger h_5^\dagger h_2 h_3 c_6
\end{align}
From the above expressions, we observe that the one-body term generated through normal-ordering acts exclusively on spin-up particles. Consequently, the shift must originate from a direct scattering process, as only such processes can produce an effective one-body potential acting solely on the spin-up sector because the normal order process is effectively doing a Hartree-Fock calculation.  The two-body interactions generated above describe interactions between  particle and hole and will contribute to the exciton energy.

As for the last term, we can group all the terms generated in the normal order process into three parts:
\begin{align}
    h_1h_2h_3h_4^\dagger h_5^\dagger h_6^\dagger =  V_{h}+ V_{hh}+ V_{hhh}
\end{align}
where $V_{h}$ represent the one-body terms acting on the hole state, $V_{2b,hh}$ represent the two-body terms between holes and the $V_{hhh}$ represent the three body term between holes.Only the one-body terms generated from the normal ordering process will contribute to the exciton energy:

\begin{align}
     V_{h}
    & = h_6^\dagger h_1 (\delta_{24}\delta_{35}-\delta_{25}\delta_{34})+h_5^\dagger h_2 (\delta_{14}\delta_{36}-\delta_{16}\delta_{34})+h_4^\dagger h_3 (\delta_{15}\delta_{26}-\delta_{16}\delta_{25})\label{eq:h one-body hartree}\\
    &+h_5^\dagger h_1(\delta_{26}\delta_{34}-\delta_{24}\delta_{36})+h_6^\dagger h_2 (\delta_{15}\delta_{34}-\delta_{14}\delta_{35})+h^\dagger_4h_2(\delta_{35}\delta_{16}-\delta_{36}\delta_{15})+h_5^\dagger h_3(\delta_{24}\delta_{16}-\delta_{15}\delta_{36})\\
    &+h_6^\dagger h_3 (\delta_{14}\delta_{25}-\delta_{15}\delta_{24} )+h_4^\dagger h_1(\delta_{36}\delta_{25}-\delta_{35}\delta_{26})
\end{align}
Note that the one body terms from the fifth, sixth and seventh terms describe direct scattering, same as the first line in Eq.~\eqref{eq:h one-body hartree}. When we consider the exciton energy, the direct term does not contribute since they canceled out between particle and hole. Therefore, we only need to consider the rest terms in $V_{1b,h}$. 
Let's list the rest one-body terms with explicitly expression:

\begin{align}
      V_{h,\text{rest}} & = -4  (\frac{1}{N_\phi}\sum_{1256}V_{5,1;1,2}V_{2,6;5,6}\frac{w(n_5)}{n_5-1})h^\dagger h - 2  (\frac{1}{N_\phi}\sum_{1256}V_{5,6;1,2}V_{1,2;5,6}\frac{w(n_5)}{n_5-1})h^\dagger h \\
     & +3 (\frac{1}{N_\phi}\sum_{1256}V_{5,1;1,2}V_{6,2;5,6}\frac{w(n_5)}{n_5-1})h^\dagger h + 3 (\frac{1}{N_\phi}\sum_{1256}V_{5,6;1,2}V_{2,1;5,6}\frac{w(n_5)}{n_5-1})h^\dagger h \\
     &\approx -0.1762 + 4.02\times 10 ^{-5} +0.1179 = -0.0583 
\end{align}
The first term in the above expression exhibits a divergence, as it corresponds to a direct scattering process that occurs only at $q = 0$. 
As for the The two-body term between particle and hole take the following form:


\begin{align}
    V_{ph} &=4\times \sum_{1,2,3,4}\sum_{6,7} V_{7,1;4,3}V_{2,6;7,6} \frac{w(n_7)}{n_7-1}c_1^\dagger h_2^\dagger h_4c_3 - 2\times \sum_{1,2,3,4}\sum_{6,7} V_{7,1;4,3}V_{6,2;7,6} \frac{w(n_7)}{n_7-1}c_1^\dagger h_2^\dagger h_4 c_3 \\
    & - 2\times \sum_{1,2,3,4}\sum_{6,7} V_{7,1;6,3}V_{2,6;7,4} \frac{w(n_7)}{n_7-1} c_1^\dagger h_2^\dagger h_4 c_3 + 2\times \sum_{1,2,3,4}\sum_{6,7} V_{7,1;6,3}V_{6,2;7,4} \frac{w(n_7)}{n_7-1}c_1^\dagger h_2^\dagger h_4 c_3 \\
    & - \sum_{1,2,3,4}\sum_{6,7} V_{1,2;7,6}V_{7,6;3,4} \frac{w(n_7)}{n_7-1}c_1^\dagger h_2^\dagger h_4 c_3  - \sum_{1,2,3,4}\sum_{6,7} V_{1,6;7,4}V_{7,2;3,6} \frac{w(n_7)}{n_7-1} c_1^\dagger h_2^\dagger h_4 c_3 
\end{align}

\begin{figure}[t]
    \centering
    \includegraphics[width=1\linewidth]{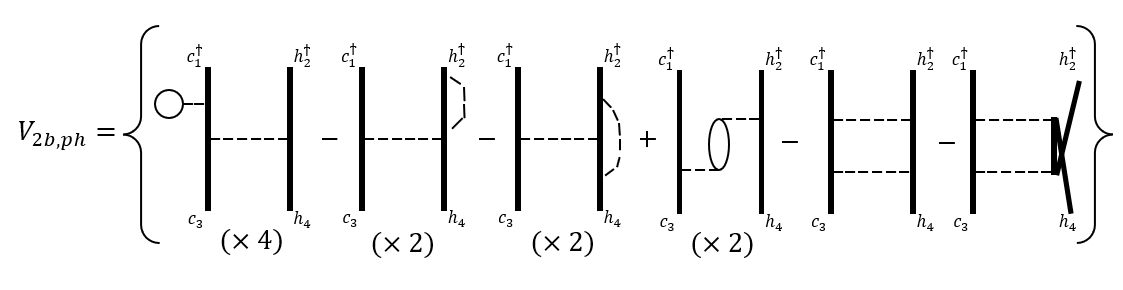}
    \caption{The diagrammatic representation of the opposite-spin(or particle-hole) two body interaction from normal ordering of three-body interaction.
 }
    \label{fig:V_2b_ph}
\end{figure}

The diagrammatic representation of the above expression is shown in 
Fig.~\ref{fig:V_2b_ph}. The first two diagrams do not contribute to the 
effective two-body interaction, since the intermediate state would require the 
same Landau-level index as the external state, which is forbidden in the 
three-body case. The remaining four diagrams correspond to the vertex correction, bubble, BCS and exchange diagram  
of the Coulomb interaction in the particle--hole transformed basis.Effectively, these four diagrams modify the original screening phase space of the effective two-body interaction.

The contribution to the exciton energy can be obtained by solving the eigenvalue equation of the two-body problem. Since the particle and hole experience the same magnetic field, the COM--relative basis block-diagonalizes $V_{ph}$, and the corresponding block eigenvalue directly contribute to the exciton energy, as the COM and relative states coincide with the spin-exciton state.

\begin{align}
    V_{ph}|M\rangle_c|m\rangle_r = E_{ph,m}|M\rangle_c|m\rangle_r
\end{align}
Together with the one-body energy shift and two-body particle-hole interaction, the three-body interaction correction to the exciton energy is given by:
\begin{align}
    W_{n=1,m} =  V_{h,rest}+E_{ph,m}
\end{align}
The results are summarized in the table below. The $n=0$ follows the same steps, so we didn't show the details here.

\begin{table}[h]
\centering
\caption{Three-body interaction contribution to the exciton energy}
\begin{tabular}{c c c c c c}
\hline
$\kappa^2$ & $m=0$ & $m=1$ & $m=2$ & $m=3$ & $m=4$ \\
\hline
$W_{n=0,m}$   & 0.0358 & 0.0972   & 0.0937   & 0.0587   & 0.0587 \\
$W_{n=1,m}$   & -0.0011   & 0.022   & 0.0048  & -0.0384   & -0.0333 \\
\hline
\end{tabular}
\end{table}


\end{document}